\documentclass{aastex}          
\usepackage{spr-astr-addons}    
\usepackage{graphicx}
\usepackage{wrapfig}
\usepackage{amsmath}
\usepackage{amssymb}
\usepackage{latexsym}
\usepackage{color}
\usepackage{subfigure}
\usepackage{enumerate}
\usepackage{textcomp}
\usepackage{hyperref}
\usepackage{bm}
\usepackage{float}
\renewcommand{\vec}[1]{\boldsymbol{#1}}

\begin{document}
%
\title{Hall-magnetohydrodynamic waves in flowing ideal
incompressible solar-wind plasmas: Reconsidered}

\shorttitle{Hall MHD waves in flowing solar-wind plasmas}
\shortauthors{Zhelyazkov et al.}

\author{I.~Zhelyazkov\altaffilmark{1*}}
\affil{Faculty of Physics, Sofia University, 1164 Sofia, Bulgaria}
\email{izh@phys.uni-sofia.bg} 
\and
\author{Z.~Dimitrov\altaffilmark{2}}
\affil{Programista JSC, 1407 Sofia, Bulgaria}
\and
\author{M.~Bogdanova\altaffilmark{1}}
\affil{Faculty of Physics, Sofia University, 1164 Sofia, Bulgaria}

\altaffiltext{1}{Faculty of Physics, Sofia University, 1164 Sofia, Bulgaria}
\altaffiltext{2}{Programista JSC, 1407 Sofia, Bulgaria}
\altaffiltext{1*}{\color{blue}izh@phys.uni-sofia.bg}

\begin{abstract}
It is well established that the magnetically structured solar atmosphere supports the propagation of  MHD waves along various kind of jets including also the solar wind.  It is well-known as well that under some conditions, namely high enough jet speeds, the propagating MHD modes can become unstable against to the most common Kelvin--Helmholtz instability (KHI).  In this article, we explore how the propagation and instability characteristics of running along a slow solar wind MHD modes are affected when they are investigated in the framework of the ideal Hall-magnetohydrodynamics.  Hall-MHD is applicable if the jet width is shorter than or comparable to the so called Hall parameter $l_\mathrm{Hall} = c/\omega_\mathrm{pi}$ (where $c$ is the speed of light and $\omega_\mathrm{pi}$ is the ion plasma frequency).  We model the solar wind as a moving with velocity $\vec{v}_0$ cylindrical flux tube of radius $a$, containing incompressible plasma with density $\rho_\mathrm{i}$ permeated by a constant magnetic field $\vec{B}_\mathrm{i}$.  The surrounding plasma is characterized with its density $\rho_\mathrm{e}$ and magnetic field $\vec{B}_\mathrm{e}$.  The dispersion relation of MHD waves is derived in the framework of both standard and Hall-MHD and is numerically solved with input parameters: the density contrast $\eta = \rho_\mathrm{e}/\rho_\mathrm{i}$, the magnetic fields ratio $b = {B}_\mathrm{e}/{B}_\mathrm{i}$, and the Hall scale parameter $l_\mathrm{Hall}/a$.  It is found that the Hall current, at moderate values of $l_\mathrm{Hall}/a$, stimulates the emerging of KHI of the kink ($m = 1)$ and high-mode ($m \geqslant 2$) MHD waves, while for the sausage wave ($m = 0$) the trend is just the opposite---the KHI is suppressed.
\end{abstract}


\section{Introduction}
\label{sec:intro}
Hall-magnetohydrodynamics (Hall-MHD) is an extend\-ed magnetohydrodynamic model in between the two-fluid theory and the standard MHD \citep{Hagstrom2014}.  That extension consists in including the Hall term, $m_\mathrm{i}(\vec{j} \times \vec{B})/(e \rho)$, in the generalized Ohm's law.   Hall-MHD describes the behavior of a plasma at length scales comparable with or shorter than an ion inertial length, $l_\mathrm{Hall} = c/\omega_\mathrm{pi}$ (where $c$ is the speed of light and $\omega_\mathrm{pi}$ is the ion plasma frequency) and time scales comparable to or shorter than the ion-cyclotron period, $\omega_\mathrm{ci}^{-1}$ \citep{Huba1995} but is simpler than the two-fluid theory because it has fewer variables.  In this way, the Hall-MHD is possible to describe waves with angular frequencies up to $\omega \approx \omega_\mathrm{ci}$.  On the other hand, since this extended model of MHD still neglects the electron mass, it is limited to angular frequencies well below the lower hybrid frequency $\omega \ll \omega_\mathrm{lh}$.  If plasmas in magnetically structured solar system are flowing, they can become unstable and one of the most important instabilities is the Kelvin--Helmholtz instability (KHI).  It occurs in the presence of a large shear flow across a thin boundary layer.  The KHI is highly efficient to mix material and momentum from both sides of a shear flow boundary.  Therefore, its macroscopic effect is equivalent to diffusion and viscosity.  \citet{Chandrasekhar1961}, in studying the KHI in incompressible homogeneous flowing plasmas on the two sides of a discontinuous tangential flow including a homogeneous magnetic filed, has shown that if the nonzero shear flow $\vec{v}_1 - \vec{v}_2$ exceeds a critical value, there rises an unstable mode whose growth rate is proportional to the wave vector of the perturbation $\vec{k}$.  It has been established also that the magnetic field can entirely stabilize the mode if there is a sufficiently large magnetic field component along the $\vec{k}$ vector.  One of the main goals of our study is to see how the Hall term in the generalized Ohm's law will change the KHI characteristics.

Hall-MHD has helped researchers progress towards the solution of a number of difficult problems, particularly magnetic reconnection in plasmas with very low resistivity.  \citet{Nykyri2004} explored the influence of the Hall term on KHI and reconnection inside KH vortices.  They have demonstrated, using a 2-D Hall-MHD simulation code, that the KHI in its nonlinear stage can develop small-scale filamentary field and current structures at the flank boundaries of the magnetosphere.   \citet{Birn2005}, investigating the magnetic reconnection in a Harris current sheet [see \citet{Harris1962}] using full particle, hybrid, and Hall-MHD simulations, obtained very similar reconnection rates, although the onset times differ, not only between different codes, but also for similar codes.  All studies also showed nearly the same final amount of reconnected flux.   \citet{Leroy2017} studied the influence of environmental parameters on mixing and reconnection caused by the KHI at the magnetopause, more specifically the different configurations than can occur in the KHI scenario in a 3-D Hall-MHD code, where the double mid-latitude reconnection process is triggered by the equatorial roll-ups.

The Hall term influences parametric instabilities of parallel propagating incoherent Alfv\'en waves \citep{Nariyuki2007}] as well as of circularly polarized small-amplitude Alfv\'en waves \citep{Ruderman2008}.  Particularly, \citet{Ruderman2008} studied the stability of circularly polarized Alfv\'en waves (pump waves) with small non-dimensional amplitude $a$ ($a \ll 1$) at $b < 1$, where $b$ is the ratio of the sound and Alfv\'en speed.  It was found that the stability properties of right-hand polarized waves are qualitatively the same as in ideal MHD.  For any values of $b$ and the dispersion parameter $\tau = k_0 v_\mathrm{A}/\omega_0$ (where $\omega_0$ and $k_0$ are the angular frequency and the wavenumber of the pump wave, respectively, and $v_\mathrm{A}$ is the Alfv\'en speed) they are subject to decay instability that occurs for wavenumbers from a band with width of order $a$.
The instability growth rate is also of order $a$.  The left-hand polarized waves can be subject to three different types of instabilities depending on the values of $b$, $a$, and $\tau$, notably modulation, decay, and beat instabilities.  In the solar wind, the effect of the Hall current generated perpendicular to the ambient magnetic field, according to \citet{Ballai2003}, can influence the plasma behavior.  More specifically, the Hall current introduces wave dispersion which may compensate the nonlinear steepening of waves.  In the presence of viscosity, these effects lead to a slowly decaying KdV soliton.  Low-frequency magnetic field fluctuations which are observed in space plasmas can steepen into very large amplitude wave phenomena, \emph{e.g.}, short large-amplitude magnetic structures, shocklets or discrete wave packets \citep{Miteva2008}, which in the case of stationary (nonlinear) waves can appear as oscillatory and solitary types solutions to the Hall-MHD equations.

Over the past decade, a great interest arose in studying Hall-MHD effects in partially ionized plasmas. \citet{Pandey2008} were the first to develop an approximate single-fluid description of a partially ionized plasma that becomes exact in the fully ionized and weakly ionized limits.  Their treatment includes the effects of ohmic, ambipolar and Hall diffusion.  These authors showed that both ambipolar and Hall diffusion depend upon the fractional ionization of the medium.  \citet{Pandey2008} found that in the ambipolar regime wave damping is dependent on both fractional ionization and ion--neutral collision frequencies, whilst in the Hall regime, where the frequency of a whistler wave is inversely proportional to the fractional ionization, and bounded by the ion--neutral collision frequency, the Hall diffusion plays an important role in the Earth's ionosphere, solar photosphere and astrophysical discs.  In the same line, \citet{Cally2015} showed that the fast-to-Alfv\'en mode conversion mediated by the Hall current depends on the ionization fraction $f$ being as low as $10^{-4}$ in the Sun.  The Hall current can couple low-frequency Alfv\'en and magnetoacoustic waves via the dimensionless Hall parameter $\epsilon = \omega/\Omega_\mathrm{i}f$, where $\omega$ is the wave angular frequency and $\Omega_\mathrm{i}$ is the mean ion gyrofrequency.  It is found, in a cold (zero-$\beta$) plasma approximation, that Hall coupling preferentially occurs where the wavevector is nearly field-aligned.  In these circumstances, Hall coupling in theory produces a continual oscillation between fast and Alfv\'en modes as the wave passes through the weakly ionized region.  At the same conditions ($f \cong 10^{-4}$), \citet{Khomenko2016} found that the ion--neutral interaction in the partially ionized solar plasma can have significant effects on the dynamical processes and on the energy balance.  She has demonstrated that neutrals can affect wave propagation, plasma instabilities (Hall instability in the presence of a flow shear, Farley--Buneman instability, as well as contact instabilities), reconnection and other fundamental processes taking place in the solar atmosphere.  Hall instability in the magnetic network consisting of vertical solar flux tubes in the presence of shear flows was recently investigated by \citet{Pandey2012} and it was ascertained that the network is dominated by Hall drift in the photosphere--lower chromosphere region (${\leqslant}1$~Mm).  In the internetwork regions with weak magnetic field, Hall drift dominates above $0.25$~Mm in the photosphere and below $2.5$~Mm in the chromosphere.  Although Hall drift does not cause any dissipation in the ambient plasma, it can destabilize the flux tubes and magnetic elements in the presence of azimuthal shear flow.  The maximum growth rate of Hall instability is proportional to the absolute value of the shear gradient, and it is dependent on the ambient diffusivity.  In a subsequent article, \citet{Pandey2013} exploring the stability of magnetic elements in the network and internetwork regions by assuming a typical shear flow gradient of ${\sim}0.1$~s$^{-1}$, showed that the magnetic diffusion shear instability grows on a time-scale of $1$~min.  Thus, it is plausible that network--internetwork magnetic elements are subject to this fast growing, diffusive shear instability, which could play an important role in driving low-frequency turbulence in the plasma in the solar photosphere and chromosphere.  Using the single-fluid description of the partially ionized plasma [see \citet{Pandey2008}], \citet{Panday2013a} studied the properties of the low-frequency surface waves in an incompressible plasma slab---the geometry is the same as in the article of \citet{Edwin1982}, namely a layer of thickness $2 x_0$ with piecewise constant density permeated by the uniform vertical magnetic field $\vec{B} = B \hat{z}$.  The thickness of the slab represents the diameter of the magnetic flux tube.  The derived wave dispersion relation [see Eq.~(49) in \citet{Panday2013a}] has rather complex form compared with the similar equation of \citet{Edwin1982}.  Note also that in deriving the dispersion relation, Panday had to use four boundary conditions [see \citet{Zhelyazkov1996}], while in the framework of the standard MHD they are only two: the continuity of the total (thermal plus magnetic pressure) and the transverse component of the Lagrangian displacement, $\xi_x$.  The KHI in partially ionized cylindrical magnetic flux tubes of dense and cool plasma surrounded by a hotter and lighter environment, was recently derived by \citet{Gomez2015}.  The authors used the governing equations of two-fluid plasma consisting of electrons and one-atom ions and the obtained wave dispersion equation [see Eq.~(23) in \citet{Gomez2015}] represents a generalization of the well-known dispersion equation of  \citet{Edwin1983}.  The main finding is the fact that the presence of a neutral component in a plasma may contribute to the onset of the KHI even for sub-Alfv\'enic longitudinal shear flows. Collisions between ions and neutrals reduce the growth rates of the unstable perturbations, but cannot completely suppress the instability.

The linear MHD wave propagation and KHI in the framework of ideal Hall-MHD were explored in 1990's and early 2000's.  The main subject in those studies were fast, kink ($m = 1$) and sausage ($m = 0$), modes traveling in plane geometries: semi-infinite or slab structures of flowing compressible or incompressible magnetized plasmas.  An extensive review of those studies plus the wave propagation in magnetically structured flux tubes in the limit of standard MHD, the reader can find in  \citet{Zhelyzkov2009} and references therein.  An attempt to generalize the parallel propagation of Hall-MHD waves in cylindrical flowing plasmas was carried out by \citet{Zhelyazkov2010}.  As a target he used the solar wind modeling it as a cylindrical flux tube with radius $a$ of homogeneous incompressible plasma with density $\rho_\mathrm{i}$ embedded in a homogeneous magnetic field $\vec{B}_0$ and moving with velocity $\vec{U}$.  The environment was assumed to be also a homogeneous medium with density $\rho_\mathrm{e}$ immersed in the same magnetic field $\vec{B}_0$.  The dispersion relation of the Hall-MHD modes was derived from the corresponding linearized governing equations of the Hall-MHD.  That derivation possesses, however, a flaw, notably the thermal pressure was ignored, which implies that the axial fluid velocity and magnetic field perturbations, $v_{1z}$ and $B_{1z}$, respectively, were set to zero.  Thus, there was obtained a rather strange model of neither incompressible nor cold plasma of the solar wind.  The Hall-MHD wave propagation and KHI in the limit of cold plasmas of the solar wind and its environment in the same geometry without any simplifications was recently studied by \citet{Zhelyazkov2018}.  A distinctive feature of the wave dispersion derived is the possibility for the existence not only of kink and sausage Hall-MHD waves, but also of high mode ($m \geqslant 2$) ones.  In this article, we reconsider the derivation of the wave dispersion relation in the limit of incompressible media for the solar wind and its surrounding plasma with keeping the thermal pressure term in the momentum equation.

The paper is organized as follows: in the next section, we discuss the geometry of the problem, equilibrium magnetic field configuration, basic physical parameters of the explored jet and present the derivation of the wave dispersion relation.  Section~3 deals with the solutions to the wave dispersion relation and their discussion.  In the last Section~4, we summarize the main findings in our research and outlook the further improvement of the new approach.


\section{Jet's geometry, governing MHD equations and dispersion relation}
\label{sec:geometry}
\begin{figure}[!ht]
 \centerline{\includegraphics[width=\columnwidth]{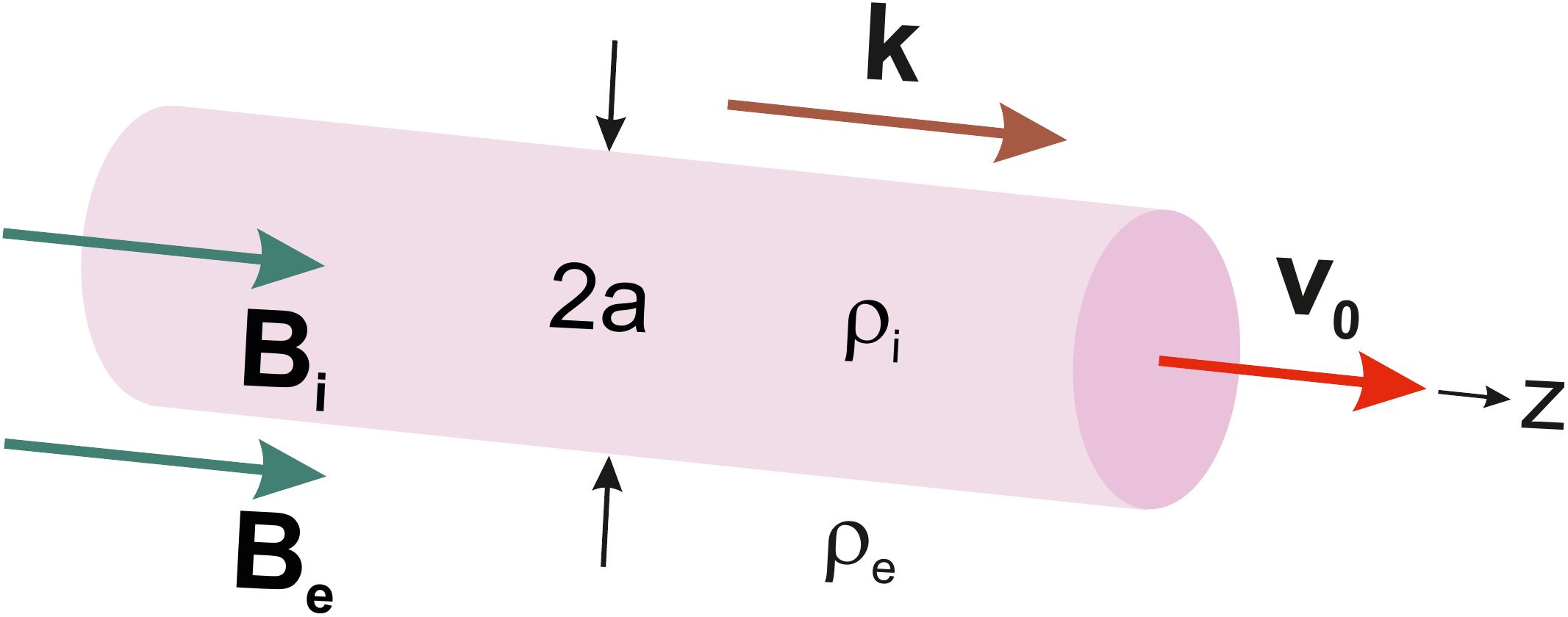}}
 \caption{Magnetic field and velocity configuration in an axially moving solar wind flux tube.}
 \label{fig:fig1}
\end{figure}
We model the solar wind as a moving with velocity $\vec{v}_0$ magnetic flux tube of radius $a$ containing incompressible plasma of homogeneous density $\rho_\mathrm{i}$ embedded in a constant magnetic field $\vec{B}_\mathrm{i}$ (see Fig.~\ref{fig:fig1}).  The environment is also an incompressible medium with homogeneous density $\rho_\mathrm{e}$ immersed in a constant magnetic field $\vec{B}_\mathrm{e}$.  Our frame of reference is attached to the environment; thus $\vec{v}_0$ should be considered as relative velocity if the surrounding plasma is flowing.  The physical parameters of the solar wind--environment configuration are: jet electron number density $n_\mathrm{i} = 2.43 \times 10^6$~m$^{-3}$ at $1$~\textsc{AU}, sound speeds $c_\mathrm{si} = c_\mathrm{se} = 70$~km\,s$^{-1}$, internal Alfv\'en speed $v_\mathrm{Ai} = 70$~km\,s$^{-1}$, external Alfv\'en speed $v_\mathrm{Ae} = 100$~km\,s$^{-1}$, and magnetic field $B_\mathrm{i} = 5$~G, which implies that $n_\mathrm{e} = 1.65 \times 10^6$~m$^{-3}$.  This value of the environment electron number density was obtained from the total pressure balance equation which states that the sum of thermal and magnetic pressure should be a constant.  Along with   $n_\mathrm{e}$, from the balance equation we can evaluate the magnetic field ratio $b \equiv B_\mathrm{e}/B_\mathrm{i} = 1.177$ and the two plasma betas, $\beta_\mathrm{i} = 1.203$ and $\beta_\mathrm{e} = 0.589$, respectively.  Thus, we have a moving jet with density contrast $\eta \equiv \rho_\mathrm{e}/\rho_\mathrm{i} = 0.679$, ion cyclotron frequency $\omega_\mathrm{ci}/2\pi = 76$~mHz, and Hall scale length (${=}v_\mathrm{Ai}/\omega_\mathrm{ci}$ which is equivalent to $c/\omega_\mathrm{pi}$) $l_\mathrm{Hall} \approx 150$~km.  This scale length is small, but not negligible compared with the tube radius of a few hundred kilometres.  Here, we introduce a scale parameter $\varepsilon = l_\mathrm{Hall}/a$ called the Hall parameter.  In the limit $\varepsilon \to 0$, the Hall-MHD system reduces to the standard MHD system.

The basic variables of the ideal Hall-MHD in the limit of incompressible media are the fluid velocity $\vec{v}(\vec{r},t)$, the magnetic field $\vec{B}(\vec{r},t)$, and the pressure $p(\vec{r},t)$.  The governing equations have the form:
\begin{equation}
\label{eq:momentum}
    \frac{\partial \vec{v}}{\partial t} + \vec{v}\cdot \nabla \vec{v} = -\frac{1}{\rho}\nabla p + \frac{1}{\rho \mu}(\nabla \times \vec{B}) \times \vec{B},
\end{equation}
\begin{equation}
\label{eq:divv}
    \nabla \cdot \vec{v} = 0,
\end{equation}
\begin{equation}
\label{eq:induction}
    \frac{\partial \vec{B}}{\partial t} = -\nabla \times \vec{E},
\end{equation}
\begin{equation}
\label{eq:Ohm'slaw}
    \vec{E} + \vec{V}_\mathrm{e} \times \vec{B} = -\frac{1}{n_\mathrm{e} e}\nabla\cdot \mathsf{P}_\mathrm{e},
\end{equation}
\begin{equation}
\label{eq:divB}
    \nabla \cdot \vec{B} = 0.
\end{equation}

Here, $\vec{E}$ is the electric field, $e$ is the elementary electric charge, $n_\mathrm{e}$ is the electron number density, and $\mu$ is the the magnetic permeability of free space.  In the generalized Ohm's law, Eq.~(\ref{eq:Ohm'slaw}) \citep{Spitzer1962}, $\vec{V}_\mathrm{e}$ is the mean velocity of electrons and $\mathsf{P}_\mathrm{e}$ is the electron pressure tensor.  We note that the normal/radial component of the magnetic field $\vec{B}$ at a tangential discontinuity (see Fig.~\ref{fig:fig1}), in equilibrium, is zero.  At this discontinuity the following three boundary conditions must be satisfied: the normal component of the Lagrangian displacement, $\xi_r$, must be continuous.  To satisfy this condition it is enough to assume that the tangential component of the bulk fluid velocity $\vec{v}$ is zero.  The total pressure (thermal plus magnetic) must be continuous, too.  We write this condition, in two-fluid approximation, as
\begin{equation}
\label{eq:2fluidtp}
    p_\mathrm{e}^\mathrm{jet} + p_\mathrm{i}^\mathrm{jet} + \frac{B_\mathrm{jet}^2}{2\mu} = p_\mathrm{e}^\mathrm{env} + p_\mathrm{i}^\mathrm{env} + \frac{B_\mathrm{env}^2}{2\mu},
\end{equation}
where here the labels `e' and `i' refer to the electrons and ions, and labels `jet' and `env' to the quantities inside and outside the discontinuity, respectively. The tangential component of the electric field, $\vec{E}_\tau$, must be continuous.  According to Eq.~(\ref{eq:Ohm'slaw}), under the assumption that the plasma is isotropic, so that $\nabla\cdot \mathsf{P}_\mathrm{e} = \nabla p_\mathrm{e}$, the expression for the electric filed has the form
\begin{equation}
\label{eq:electricfield}
    \vec{E} = -\vec{V}_\mathrm{e} \times \vec{B} - \frac{1}{n_\mathrm{e} e}\nabla p_\mathrm{e},
\end{equation}
where the mean velocity of the electrons is
\begin{equation}
\label{eq:electronvel}
    \vec{V}_\mathrm{e} = \vec{v} - \frac{\vec{j}}{n_\mathrm{e} e}.
\end{equation}
Substituting Eq.~(\ref{eq:electronvel}) in Eq.~(\ref{eq:electricfield}) yields
\begin{equation}
\label{eq:newelctricfield}
    \vec{E} = -\vec{v} \times \vec{B} + \frac{1}{n_\mathrm{e} e}\vec{j}\times \vec{B} - \frac{1}{n_\mathrm{e} e}\nabla p_\mathrm{e}.
\end{equation}
Since the normal components of $\vec{v}$ and $\vec{B}$ are equal to zero, the tangential component of the first term on the right-hand side of this equation also is zero.  We assume that all quantities are constant inside and outside the discontinuity.  Then it follows that the normal (that is, radial) component of the current density, $\vec{j} = \nabla \times \vec{B}/\mu$, is zero and, consequently, the tangential component of the second term on the right-hand side of Eq.~(\ref{eq:newelctricfield}) is zero.  The same is true for the third term, so $\vec{E}_\tau = 0$ at both sides of the discontinuity and the third boundary condition is satisfied.

However, it follows from the condition that $\vec{E}$ is finite everywhere that the electron pressure $p_\mathrm{e}$ must be continuous. Otherwise there would be a term proportional to the Dirac delta-function in the expression for $\vec{E}$.  Hence, Eq.~(\ref{eq:2fluidtp}) reduces to
\begin{equation}
\label{eq:finalpressbaleq}
    p_\mathrm{i}^\mathrm{jet} + \frac{B_\mathrm{jet}^2}{2\mu} = p_\mathrm{i}^\mathrm{env} + \frac{B_\mathrm{env}^2}{2\mu}.
\end{equation}

The basic Maxwell law of induction, Eq.~(\ref{eq:induction}), on using Eq.~(\ref{eq:electricfield}) for the electric field $\vec{E}$, read as
\begin{eqnarray}
\label{eq:induct-modified}
    \frac{\partial \vec{B}}{\partial t} = -\nabla \times \vec{E} = \nabla \times \left( \vec{V}_\mathrm{e} \times \vec{B} + \frac{\nabla p_\mathrm{e}}{n_\mathrm{e} e} \right) \nonumber \\
    \nonumber \\
    {}= \nabla \times ( \vec{V}_\mathrm{e} \times \vec{B} ).
\end{eqnarray}
Here, the $\nabla \times (\nabla p_\mathrm{e})/n_\mathrm{e}$ has been put equal to zero on the ground that the electron pressure will fluctuate in a direct functional relationship to the density and so to $n_\mathrm{e}$.  The standard equation (\ref{eq:induction}) has the well-known interpretation that the magnetic lines of force `move with,' or `are frozen into,' the gas.  The more accurate equation (\ref{eq:induct-modified}) means that they move with, or are frozen into, the \emph{electron\/} gas \citep{Lighthill1960}.  Bearing in mind that the mean velocity of electrons according to Eq.~(\ref{eq:electronvel}) is equal to
\[
    \vec{V}_\mathrm{e} = \vec{v} - \frac{\vec{j}}{n_\mathrm{e} e},
\]
after substituting it in Eq.~(\ref{eq:induct-modified}) and using the Amp\`{e}re's law, $\vec{j} = (\nabla \times \vec{B})/\mu$, we get the final form of the induction equation:
\begin{equation}
\label{eq:induction-new}
    \frac{\partial \vec{B}}{\partial t} = \nabla \times (\vec{v} \times \vec{B}) - \frac{m_\mathrm{i}}{\rho e}\nabla \times\left( \frac{1}{\mu}(\nabla \times \vec{B}) \times \vec{B} \right),
\end{equation}
in which $m_\mathrm{i}$ is the ion mass.

Equations (\ref{eq:momentum}), (\ref{eq:divv}), (\ref{eq:induction-new}), and (\ref{eq:divB}) constitute a closed set of equations.  It will be used in studying the wave propagation in the frame of Hall-MHD.

In cylindrical geometry (see Fig.~\ref{fig:fig1}), excited MHD waves propagate along the flux tube, which implies a wavevector $\vec{k} = (0, 0, k_z)$.  Considering small perturbations from equilibrium in the form
\[
    \vec{B} = \vec{B}_0 + \vec{B}_1, \; \vec{v} = \vec{v}_0 + \vec{v}_1, \; \mbox{and} \; p = p_0 + p_1,
\]
where $\vec{B}_0 = (0, 0, B_0)$ ($B_0$ being $B_\mathrm{i}$ or $B_\mathrm{e}$), $\vec{B}_1 = (B_{1r}, B_{1\phi}, B_{1z})$, $\vec{v}_0 = (0, 0, v_0)$, and $\vec{v}_1 = (v_{1r}, v_{1\phi}, v_{1z})$, the set of linearized equations which govern the dynamics of aforementioned magnetic field, velocity, and pressure perturbations in the incompressible-plasma approximation has the form
\begin{equation}
\label{eq:newmom}
    \rho \left( \frac{\partial}{\partial t} + \vec{v}_0 \cdot \nabla \right)\vec{v}_1 = -\nabla p_1 + \frac{1}{\mu}(\nabla \times \vec{B}_1 ) \times \vec{B}_0,
\end{equation}
\begin{eqnarray}
\label{eq:newind}
    \frac{\partial \vec{B}_1}{\partial t} = \nabla \times (\vec{v}_0 \times \vec{B}_1) + \nabla \times (\vec{v}_1 \times \vec{B}_0) \nonumber \\
    \nonumber \\
    {}- \frac{m_\mathrm{i}}{e \rho \mu}\nabla \times [ (\nabla \times \vec{B}_1) \times \vec{B}_0 ],
\end{eqnarray}
and the constraints
\begin{equation}
\label{eq:divv1}
    \nabla \cdot \vec{v}_1 = 0,
\end{equation}
\begin{equation}
\label{eq:divB1}
    \nabla \cdot \vec{B}_1 = 0.
\end{equation}

To investigate the stability of the jet--environment system, Eqs.~(\ref{eq:newmom})--(\ref{eq:divB1}) are Fourier transformed, assuming that all perturbations have the form
\[
    g(r,\phi,z,t) = g(r)\exp\left[ \mathrm{i}(-\omega t + m\phi + k_z z) \right],
\]
where $g$ represents any quantities $\vec{v}_1$, $\vec{B}_1$, and $p_1$; $\omega$ is the angular wave frequency, $m$ is the azimuthal mode number, and $k_z$ is the axial wavenumber.  Thus, the set of equations which govern the time and space evolution of all the perturbations has the form
\begin{equation}
\label{eq:momr}
    -\mathrm{i}\Omega v_{1r} + \frac{1}{\rho}\frac{\mathrm{d}}{\mathrm{d}r}p_\mathrm{tot} - \frac{B_0}{\mu \rho}\mathrm{i}k_z B_{1r} = 0,
\end{equation}
\begin{equation}
\label{eq:momphi}
    -\mathrm{i}\Omega v_{1\phi} + \mathrm{i} \frac{1}{\rho}\frac{m}{r} v_{1\phi} - \frac{B_0}{\mu \rho}\mathrm{i}k_z B_{1\phi} = 0,
\end{equation}
\begin{equation}
\label{eq:momz}
    -\mathrm{i}\Omega v_{1z} + \mathrm{i} \frac{1}{\rho}k_z v_{1z} - \frac{B_0}{\mu \rho}\mathrm{i}k_z B_{1z} = 0,
\end{equation}
\begin{eqnarray}
\label{eq:undr}
    -\mathrm{i}\Omega B_{1r}- \mathrm{i}k_z B_0 v_{1r} + \frac{k_z v_\mathrm{A}^2}{\omega_\mathrm{ci}} \nonumber \\
    \nonumber \\
    {}\times \left( -\frac{m}{r}B_{1z} + k_z B_{1\phi} \right) = 0,
\end{eqnarray}
\begin{eqnarray}
\label{eq:undphi}
    -\mathrm{i}\Omega B_{1\phi}- \mathrm{i}k_z B_0 v_{1\phi} - \frac{k_z v_\mathrm{A}^2}{\omega_\mathrm{ci}} \nonumber \\
    \nonumber \\
    {}\times \left( k_z B_{1r} + \mathrm{i}\frac{\mathrm{d}}{\mathrm{d}r} B_{1z} \right) = 0,
\end{eqnarray}
\begin{eqnarray}
\label{eq:undz}
    -\mathrm{i}\Omega B_{1z}- \mathrm{i}k_z B_0 v_{1z} + \mathrm{i}\frac{k_z v_\mathrm{A}^2}{\omega_\mathrm{ci}} \nonumber \\
    \nonumber \\
    {}\times \left( \frac{\mathrm{d}}{\mathrm{d}r} B_{1\phi} + \frac{1}{r} B_{1\phi} - \mathrm{i}\frac{m}{r}B_{1r} \right) = 0,
\end{eqnarray}
\begin{equation}
\label{eq:dvv}
    \left( \frac{\mathrm{d}}{\mathrm{d}r} + \frac{1}{r} \right)v_{1r} + \mathrm{i}\frac{m}{r}v_{1\phi} + \mathrm{i} k_z v_{1z} = 0,
\end{equation}
\begin{equation}
\label{eq:dvb}
    \left( \frac{\mathrm{d}}{\mathrm{d}r} + \frac{1}{r} \right)B_{1r} + \mathrm{i}\frac{m}{r}B_{1\phi} + \mathrm{i} k_z B_{1z} = 0.
\end{equation}
Here, $\Omega = \omega - k_z v_0$ is the Doppler shifted angular frequency, $v_\mathrm{A} = B_0/\sqrt{\mu \rho}$, and $p_\mathrm{tot} = p_1 + B_0 B_{1z}/\mu$ is the perturbation of the total magnetic pressure (thermal plus magnetic).

From Eq.~(\ref{eq:momr}), we obtain
\begin{equation}
\label{eq:B1r}
    B_{1r} = -\frac{\Omega}{k_z v_\mathrm{A}^2}B_0 v_{1r} - \mathrm{i}\frac{1}{k_z} \frac{\mathrm{d}}{\mathrm{d}r}\frac{\mu}{B_0}p_\mathrm{tot}.
\end{equation}
Similarly, from Eqs.~(\ref{eq:momphi}) and (\ref{eq:momz}) one finds
\begin{equation}
\label{eq:B1phi}
    B_{1\phi} = -\frac{\Omega}{k_z v_\mathrm{A}^2}B_0 v_{1\phi} + \frac{1}{k_z} \frac{m}{r}\frac{\mu}{B_0}p_\mathrm{tot},
\end{equation}
\begin{equation}
\label{eq:B1z}
    B_{1z} = -\frac{\Omega}{k_z v_\mathrm{A}^2}B_0 v_{1z} + \frac{\mu}{B_0}p_\mathrm{tot}.
\end{equation}
Above three momentum equations can be also solved to yield the fluid velocity perturbations as functions of the magnetic filed perturbations and $p_\mathrm{tot}$:
\begin{equation}
\label{eq:v1r}
    v_{1r} = -\frac{1}{\Omega}\left( \mathrm{i}\frac{1}{\rho}\frac{\mathrm{d}}{\mathrm{d}r} p_\mathrm{tot} + \frac{k_z v_\mathrm{A}^2}{B_0}B_{1r} \right),
\end{equation}
\begin{equation}
\label{eq:v1phi}
    v_{1\phi} = \frac{1}{\Omega}\left( \frac{1}{\rho}\frac{m}{r} p_\mathrm{tot} - \frac{k_z v_\mathrm{A}^2}{B_0}B_{1\phi} \right),
\end{equation}
\begin{equation}
\label{eq:v1z}
    v_{1z} = \frac{1}{\Omega}\left( \frac{1}{\rho}k_z p_\mathrm{tot} - \frac{k_z v_\mathrm{A}^2}{B_0}B_{1z} \right).
\end{equation}
Further on, from Eq.~(\ref{eq:undr}) we obtain
\[
    v_{1r} = -\frac{1}{k_z B_0}\left[ \Omega B_{1r} + \mathrm{i}\frac{k_z v_\mathrm{A}^2}{\omega_\mathrm{ci}}\left( -\frac{m}{r}B_{1z} + k_z B_{1\phi} \right) \right].
\]
Now we substitute this $v_{1r}$ into the expression for $B_{1r}$ given by Eq.~(\ref{eq:B1r}) to find
\[
    B_{1r} = \mathrm{i}H \frac{\mathrm{d}}{\mathrm{d}r} p_\mathrm{tot} + \mathrm{i} \frac{\epsilon}{C - 1}\left( \frac{1}{k_z}\frac{m}{r}B_{1z} - B_{1\phi} \right).
\]
In a similar way, we obtain
\[
    B_{1\phi} = -H \frac{m}{r} p_\mathrm{tot} + \frac{\epsilon}{C - 1}\left( \mathrm{i} B_{1r} - \frac{1}{k_z}\frac{m}{r}B_{1z} \right),
\]
\begin{eqnarray*}
    B_{1z} = -H k_z p_\mathrm{tot} + \frac{\epsilon}{C - 1} \nonumber \\
    \nonumber \\
    {}\times\left[ \frac{1}{k_z}\left( \frac{\mathrm{d}}{\mathrm{d}r} + \frac{1}{r} \right)B_{1\phi} - \mathrm{i}\frac{1}{k_z} \frac{m}{r}B_{1r} \right],
\end{eqnarray*}
where
\begin{equation}
\label{eq:eps}
    H = \frac{1}{\rho} \frac{k_z B_0}{\Omega^2 - k_z^2 v_\mathrm{A}^2}, \; \epsilon = \frac{\Omega}{\omega_\mathrm{ci}}, \; \mbox{and} \; C = \left( \frac{\Omega}{k_z v_\mathrm{A}} \right)^2.
\end{equation}
After substituting these expressions of $B_{1r}$, $B_{1\phi}$, and $B_{1z}$ into Eq.~(\ref{eq:divB}) we derive a second order ordinary differential equation for $p_\mathrm{tot}$, namely
\begin{equation}
\label{eq:Bessel}
    \left[ \frac{\mathrm{d}^2}{\mathrm{d}r^2} + \frac{1}{r}\frac{\mathrm{d}}{\mathrm{d}r} - \left(  k_z^2 + \frac{m^2}{r^2} \right) \right]p_\mathrm{tot} = 0.
\end{equation}
This is the Bessel equation for the modified Bessel function of second kind and the solutions to it in the two media are
\begin{equation}
\label{eq:ptot}
    p_{\mathrm{tot}}(r) = \left\{ \begin{array}{lc}
                                 \alpha_\mathrm{i}I_m(k_z r) & \mbox{for  $\;\,r \leqslant a$} \\
                                 \alpha_\mathrm{e}K_m(k_z r) & \mbox{for  $\;r > a$},
                                 \end{array}
                         \right.
\end{equation}
where $\alpha_\mathrm{i}$ and $\alpha_\mathrm{e}$ are constants.

The next step is to express $v_{1r}$ in terms of $p_\mathrm{tot}$.  From Eq.~(\ref{eq:undr}), using the expressions (\ref{eq:B1phi}) and (\ref{eq:B1z}), we obtain
\begin{equation}
\label{eq:newB1r}
    B_{1r} = -k_z B_0 \left[ \frac{1}{\Omega}v_{1r} - \mathrm{i}\frac{1}{\omega_\mathrm{ci}}v_{1\phi} + \mathrm{i}\frac{1}{\omega_\mathrm{ci}}\frac{1}{k_z}\frac{m}{r}v_{1z} \right].
\end{equation}
Similarly, from Eqs.~(\ref{eq:undphi}) and (\ref{eq:undz}) we find
\begin{equation}
\label{eq:newB1phi}
    B_{1\phi} = -k_z B_0 \left[ \frac{1}{\Omega}v_{1\phi} + \mathrm{i}\frac{1}{\omega_\mathrm{ci}}v_{1r} - \frac{1}{\omega_\mathrm{ci}}\frac{1}{k_z}\frac{\mathrm{d}}{\mathrm{d}r}v_{1z} \right],
\end{equation}
\begin{eqnarray}
\label{eq:newB1z}
    B_{1z} = -k_z B_0 \left[ \frac{1}{\Omega}v_{1z} + \frac{1}{\omega_\mathrm{ci}}\frac{1}{k_z} \left( \frac{\mathrm{d}}{\mathrm{d}r} + \frac{1}{r} \right) v_{1\phi} \right. \nonumber \\
    \nonumber \\
    \left.
    {}- \mathrm{i}\frac{1}{\omega_\mathrm{ci}}\frac{1}{k_z}\frac{m}{r}v_{1r} \right].
\end{eqnarray}
In the above two equations for ${B}_{1r}$ and ${B}_{1\phi}$, we replace the $z$ component of fluid velocity perturbation, $\vec{v}_{1z}$, by its value obtained from momentum equation (\ref{eq:momz}), notably
\[
    v_{1z} = \frac{1}{\rho \Omega}k_z p_1,
\]
and obtain that
\[
    B_{1r} = -k_z B_0 \left[ \frac{1}{\Omega}v_{1r} - \mathrm{i}\frac{1}{\omega_\mathrm{ci}}v_{1\phi} + \mathrm{i} \frac{\Omega}{\omega_\mathrm{ci}}\frac{1}{\rho \Omega^2} \frac{m}{r}p_1 \right],
\]
\[
    B_{1\phi} = -k_z B_0 \left[ \frac{1}{\Omega}v_{1\phi} + \mathrm{i}\frac{1}{\omega_\mathrm{ci}}v_{1r} - \mathrm{i} \frac{\Omega}{\omega_\mathrm{ci}}\frac{1}{\rho \Omega^2} \frac{\mathrm{d}}{\mathrm{d}r}p_1 \right].
\]
Note that the third members in the brackets of above new expressions of ${B}_{1r}$ and ${B}_{1\phi}$ are multiplied by the small parameter of the problem, $\epsilon = \Omega/\omega_\mathrm{ci}$.  In such a case we can ignore them---otherwise the problem becomes not tractable analytically.  In other words, the expressions of $B_{1r}$ and $B_{1\phi}$ which we will use have the form
\[
    B_{1r} = -k_z B_0 \left[ \frac{1}{\Omega}v_{1r} - \mathrm{i}\frac{1}{\omega_\mathrm{ci}}v_{1\phi} \right],
\]
\[
    B_{1\phi} = -k_z B_0 \left[ \frac{1}{\Omega}v_{1\phi} + \mathrm{i}\frac{1}{\omega_\mathrm{ci}}v_{1r} \right].
\]
Now we insert the above $B_{1r}$ into the expression of $v_{1r}$ given by Eq.~(\ref{eq:v1r}) and obtain
\[
    v_{1r} = \mathrm{i}\frac{\Omega}{k_z^2 v_\mathrm{A}^2 - \Omega^2}\left( \frac{1}{\rho} \frac{\mathrm{d}}{\mathrm{d}r} p_\mathrm{tot} + \frac{k_z^2 v_\mathrm{A}^2}{\omega_\mathrm{ci}}  v_{1\phi} \right).
\]
Similarly, inserting the latest expression of $B_{1\phi}$ into Eq.~(\ref{eq:v1phi}) we find
\[
    v_{1\phi} = -\frac{\Omega}{k_z^2 v_\mathrm{A}^2 - \Omega^2}\left( \frac{1}{\rho} \frac{m}{r}p_\mathrm{tot} + \mathrm{i}\frac{k_z^2 v_\mathrm{A}^2}{\omega_\mathrm{ci}} v_{1r} \right).
\]
The final step is to replace $v_{1\phi}$ in the above expression of $v_{1r}$ to obtain the radial component of fluid velocity perturbation in terms of $p_\mathrm{tot}$ and its derivative $\mathrm{d} p_\mathrm{tot}/\mathrm{d}r$, namely
\begin{eqnarray*}
    \left[ 1 - \left( \frac{k_z^2 v_\mathrm{A}^2}{k_z^2 v_\mathrm{A}^2 - \Omega^2} \right)^2 \frac{\Omega^2}{\omega_\mathrm{ci}^2}\right]v_{1r} = \mathrm{i}\frac{\Omega}{k_z^2 v_\mathrm{A}^2 - \Omega^2}\frac{1}{\rho} \nonumber \\
    \nonumber \\
    {}\times\left( \frac{\mathrm{d}}{\mathrm{d}r}  - \frac{k_z^2 v_\mathrm{A}^2}{k_z^2 v_\mathrm{A}^2 - \Omega^2}\frac{\Omega}{\omega_\mathrm{ci}}\frac{m}{r} \right)p_\mathrm{tot}.
\end{eqnarray*}
The coefficient in front of $v_{1r}$ takes the form
\begin{eqnarray*}
    1 - \left( \frac{k_z^2 v_\mathrm{A}^2}{k_z^2 v_\mathrm{A}^2 - \Omega^2} \right)^2 \epsilon^2 = 1 - \left( \frac{1}{1 - C} \right)^2 \epsilon^2 \nonumber \\
    \nonumber \\
    {}= 1 - \left( \frac{\epsilon}{1 - C} \right)^2.
\end{eqnarray*}
Hence,
\begin{equation}
\label{eq:virfinal}
    v_{1r} = \mathrm{i}\frac{1}{\rho}\frac{\Omega}{k_z^2 v_\mathrm{A}^2 - \Omega^2}\frac{1}{Z} \left( \frac{\mathrm{d}}{\mathrm{d}r} - \frac{\epsilon}{1 - C}\frac{m}{r}  \right)p_\mathrm{tot},
\end{equation}
where $Z = 1 - \epsilon^2/(1 - C)^2$.

Thus, the radial component (\ref{eq:virfinal}) of the velocity perturbation has the following presentations in both media:
\begin{eqnarray*}
    v_{1r}(r \leqslant a) = -\mathrm{i}\frac{1}{\rho_\mathrm{i}}\frac{\omega - k_z v_0}{( \omega - k_z v_0 )^2 - k_z^2 v_\mathrm{Ai}^2}\frac{\alpha_\mathrm{i}}{Z_\mathrm{i}} \nonumber \\
    \nonumber \\
    {}\times\left[ k_z I_m^\prime(k_z r) - \frac{\epsilon_\mathrm{i}}{1 - C_\mathrm{i}}\frac{m}{r} I_m(k_z r) \right],
\end{eqnarray*}
\begin{eqnarray*}
    v_{1r}(r > a) = -\mathrm{i}\frac{1}{\rho_\mathrm{e}}\frac{\omega}{\omega^2 - k_z^2 v_\mathrm{Ae}^2}\frac{\alpha_\mathrm{e}}{Z_\mathrm{e}} \nonumber \\
    \nonumber \\
    {}\times\left[ k_z K_m^\prime(k_z r) - \frac{\epsilon_\mathrm{e}}{1 - C_\mathrm{e}}\frac{m}{r} K_m(k_z r) \right],
\end{eqnarray*}
where the prime means differentiation of the Bessel function on its argument.  Finally, by
applying the boundary conditions for continuity of the ratio $v_{1r}/\Omega$ [\emph{aka\/} the radial component of the Lagrangian displacement $\boldsymbol{\xi}$ \citep{Chandrasekhar1961}] and the total pressure perturbation $p_\mathrm{tot}$ at $r = a$, we obtain the dispersion relation of normal MHD modes propagating on a moving incompressible-plasma magnetic flux tube
\begin{eqnarray}
\label{eq:dispeqn}
    \frac{\rho_\mathrm{e}}{\rho_\mathrm{i}}\left( \omega^2 - k_z^2 v_\mathrm{Ae}^2 \right)Z_\mathrm{e}\left( k_z \frac{I_m^{\prime}(k_z a)}{I_m(k_z a)} - \frac{\varepsilon_\mathrm{i}}{1 - C_\mathrm{i}} \frac{m}{a} \right) \nonumber \\
    \nonumber \\
    {}-\left[ \left( \omega - k_z v_0 \right)^2 - k_z^2 v_\mathrm{Ai}^2 \right] \nonumber \\
    \nonumber \\
    {}\times Z_\mathrm{i}\left( k_z \frac{K_m^{\prime}(k_z a)}{K_m(k_z a)} - \frac{\varepsilon_\mathrm{e}}{1 - C_\mathrm{e}} \frac{m}{a} \right) = 0.
\end{eqnarray}
When $\epsilon_\mathrm{i} = \epsilon_\mathrm{e} = 0$, one obtains the well-known dispersion relation of the MHD normal modes propagating along a flowing incompressible plasma.

\section{Numerical results and discussion}
\label{sec:numerics}
Let us recall that dispersion relation~(\ref{eq:dispeqn}) is applicable when both media, the jet and its environment, are incompressible plasmas.  In Section~2, we have evaluated the plasma betas of the solar wind and its surrounding plasma as $\beta_\mathrm{i} = 1.203$ and $\beta_\mathrm{e} = 0.589$, respectively.  In this case, it is more natural to treat the jet environment as a cool medium with zero beta.  It is instructive to explore how such a modification will change the propagation and instability characteristics of a given MHD mode, say the kink ($m = 1$) one.  It is better the comparison to be made in the limit of the standard MHD, because we can solve numerically the wave dispersion relation considering both media as compressible plasmas, as well as treating the jet as an incompressible medium and its environment as a cool plasma.  To this end, we display here the well-known wave dispersion equation of the normal MHD modes of flowing compressible magnetized plasmas
\citep{Homen2003,Nakariakov2007,Zhelyazkov2012}:
\begin{eqnarray}
\label{eq:dispeqcom}
        \frac{\rho_{\rm e}}{\rho_{\rm i}}\left( \omega^2 - k_z^2 v_{\rm Ae}^2
        \right) m_{0{\rm i}}\frac{I_m^{\prime}(m_{0{\rm
        i}}a)}{I_m(m_{0{\rm i}}a)} \nonumber \\
        \nonumber \\
        {}- \left[ \left( \omega - k_z
        v_0 \right)^2 - k_z^2 v_{\rm Ai}^2 \right] \nonumber \\
        \nonumber \\
        {}\times m_{0{\rm e}}\frac{K_m^{\prime}(m_{0{\rm e}}a)}{K_m(m_{0{\rm e}}a)} = 0,
\end{eqnarray}
where the squared wave attenuation coefficients in both media are given by the expression
\[
    m_0^2 = -\frac{\left( \Omega^2 - k_z^2 c_{\rm s}^2 \right)\left( \Omega^2 - k_z^2 v_{\rm A}^2 \right)}{\left( c_{\rm s}^2 + v_{\rm A}^2 \right)\left( \Omega^2 - \omega_{\rm c}^2 \right)},
\]
in which $\Omega \equiv \omega$ in the environment, and the cusp frequency, $\omega_\mathrm{c}$, is usually expressed via the so-called tube speed, $c_\mathrm{T}$, namely $\omega_\mathrm{c} = k_z  c_\mathrm{T}$, where \citep{Edwin1983}
\[
    c_\mathrm{T} = \frac{c_{\rm s}v_{\rm A}}{\sqrt{c_{\rm s}^2 + v_{\rm A}^2}}.
\]
We recall that for the kink mode ($m = 1$) one defines the so-called kink speed \citep{Edwin1983},
\begin{equation}
\label{eq:kinkspeed}
        c_{\rm k} = \left( \frac{\rho_{\rm i} v_{\rm Ai}^2 + \rho_{\rm e}
        v_{\rm Ae}^2}{\rho_{\rm i} + \rho_{\rm e}} \right)^{1/2},
\end{equation}
which, as seen, is independent of sound speeds and characterizes the propagation of transverse perturbations.  We will show, that the kink mode can become unstable against the KH instability.

In the case of incompressible jet--cool environment configuration the above dispersion equation takes the form:
\begin{eqnarray}
\label{eq:dispeqmix}
        \frac{\rho_{\rm e}}{\rho_{\rm i}}\left( \omega^2 - k_z^2 v_{\rm Ae}^2
        \right) k_z\frac{I_m^{\prime}(k_z a)}{I_m(k_z a)} \nonumber \\
        \nonumber \\
        {}- \left[ \left( \omega - k_z
        v_0 \right)^2 - k_z^2 v_{\rm Ai}^2 \right]\nonumber \\
        \nonumber \\
        {}\times  m_\mathrm{0e}^\mathrm{c}\frac{K_m^{\prime} (m_\mathrm{0e}^\mathrm{c}a)}{K_m(m_\mathrm{0e}^\mathrm{c}a)} = 0,
\end{eqnarray}
where now the wave attenuation coefficient inside the jet is simply equal to $k_z$, while that in the cool environment is $m_\mathrm{0e}^\mathrm{c} = k_z \left( 1 - \omega^2/k_z^2 v_\mathrm{Ae}^2 \right)^{1/2}$.

Since we are looking for unstable MHD modes, we consider the wave angular frequency, $\omega$, as a complex quantity: $\omega = \mathrm{Re}(\omega) + \mathrm{i}\mathrm{Im}(\omega)$, but the mode number $m$ and the axial wavenumber $k_z$ are assumed to be real quantities.  The numerical solutions to all dispersion relations will be carried out in non-dimensional variables.  Thus, we normalize all velocities with respect to the Alfv\'en speed inside the flux tube, $v_\mathrm{Ai}$, and the wavelength $\lambda = 2\pi/k_z$ with respect to the tube radius $a$, which implies that we shall look for solutions of the normalized complex wave phase velocity $\omega/k_z v_\mathrm{Ai}$ as a function of the normalized wavenumber $k_z a$ and input parameters, whose number depends upon the form of the wave dispersion relation.  When we explore Eq.~(\ref{eq:dispeqcom}), along with the density contrast $\eta$, we should evaluate the reduced plasma betas, $\tilde{\beta}_\mathrm{i,e} = c_\mathrm{i,e}^2/v_\mathrm{Ai,e}^2$, magnetic field ratio, $b = B_\mathrm{e}/B_\mathrm{i}$, and Alfv\'en Mach number $M_\mathrm{A} = v_0/v_\mathrm{Ai}$, which represents the flow velocity ${v}_0$.  We note that for normalization of the sound speeds one needs the parameters $\tilde{\beta}_\mathrm{i,e}$, while $b$ is used at the normalization of the Alfv\'en speed in the environment, $v_\mathrm{Ae}$.

At given sound and Alfv\'en speeds, alongside the input parameters $\eta$ and $b$, one can make some predictions, namely to evaluate the value of the kink speed (\ref{eq:kinkspeed}) in a static flux tube, and the expected threshold/critical Alfv\'en Mach number at which the KHI would start---the latter is determined by the inequality \citep{Zaqarashvili2014}
\begin{equation}
\label{eq:criterion}
    |m|M_\mathrm{A}^2 > (1 + 1/\eta)(|m|b^2 + 1).
\end{equation}

\begin{figure}[!ht]
  \centering
\subfigure{\includegraphics[width = 3.3in]{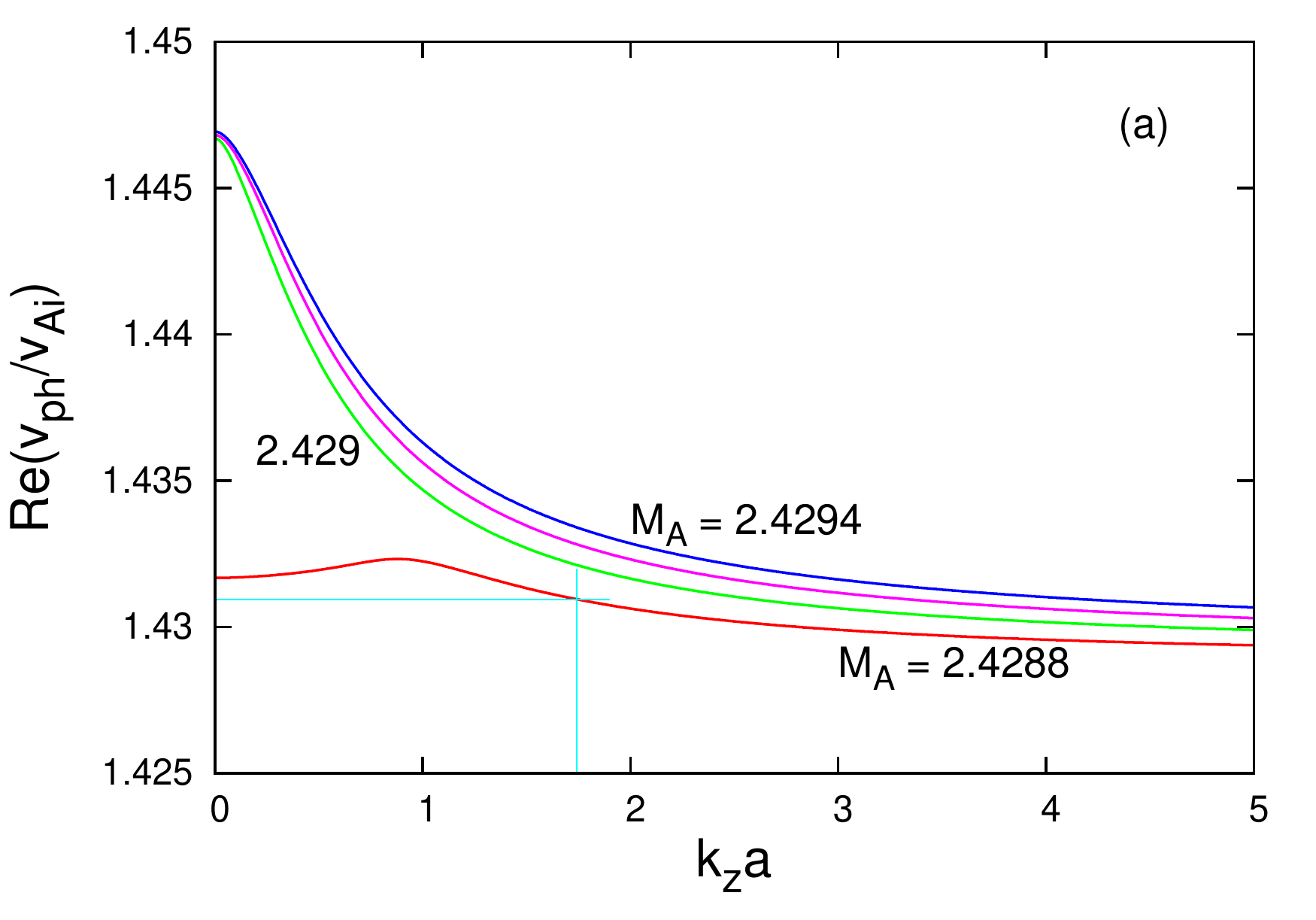}} \\
\subfigure{\includegraphics[width = 3.3in]{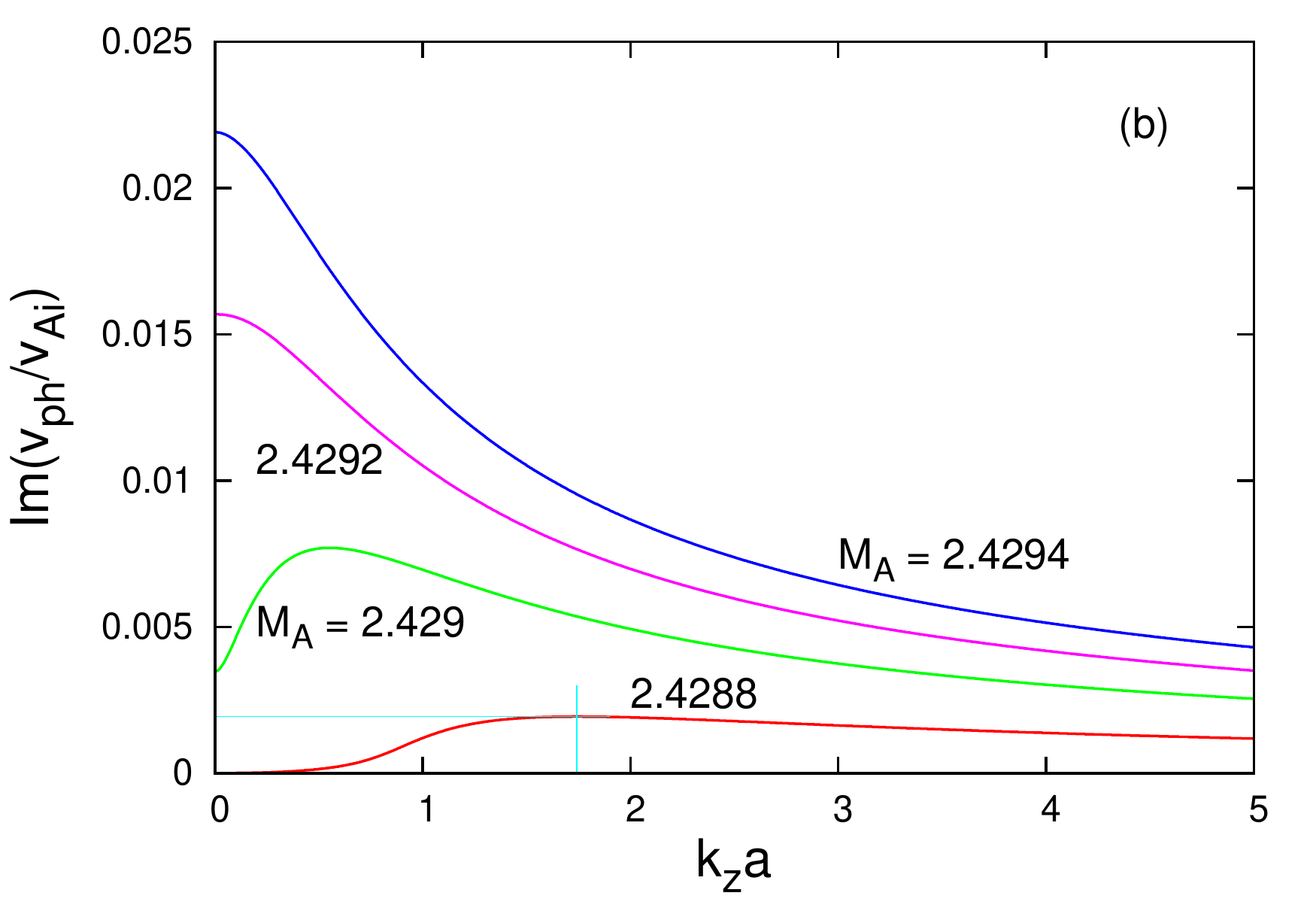}}
  \caption{\textbf{(a)} Dispersion curves of the kink mode ($m = 1$) propagating along a compressible flowing magnetic flux tube at $\eta = 0.679$, $b = 1.177$, $\tilde{\beta}_\mathrm{i} = 1.002$, and $\tilde{\beta}_\mathrm{e} = 0.491$ for four Alfv\'en Mach numbers equal to $2.4288$, $2.429$, $2.4292$, and $2.4294$, respectively.  \textbf{(b)} Growth rates of of the unstable kink mode at the same input parameters.  The marginally dispersion and growth rate curves are obtained at $M_\mathrm{A} = 2.4288$ and are plotted in red color.  The crosses of cyan lines yield the normalized values of the wave growth rate and wave angular frequency at $k_z a = 1.743$---the maximum of the growth rate curve.}
   \label{fig:fig2}
\end{figure}

Let us first start with finding the solutions to dispersion relation~(\ref{eq:dispeqcom}).  The input parameters of the numerical task are: $m = 1$, $\eta = 0.679$, $b = 1.177$, $\tilde{\beta}_\mathrm{i} = 1.002$, and $\tilde{\beta}_\mathrm{e} = 0.491$.  Alfv\'en Mach number, $M_\mathrm{A}$, is a running parameter. According to the criterion~(\ref{eq:criterion}), one can expect the emergence of KHI instability at $M_\mathrm{A} \geqslant 2.429$.  The normalized kink speed~(\ref{eq:kinkspeed}) in an immobile flux tube ($M_\mathrm{A} = 0$) is $c_\mathrm{k}/v_\mathrm{Ai} = 1.192$.  The results of numerical calculations are presented in Fig.~\ref{fig:fig2}.
The first and the most impressive result is the fact that the threshold Alfv\'en Mach number at which the KHI occurs is equal to $2.4288$---a value very close to the predicted one.  With $v_\mathrm{Ai} = 70$~km\,s$^{-1}$ this implies a critical flow velocity of $170.0$~km\,s$^{-1}$, which is accessible for the slow solar wind with speeds less than or equal to $350.0$~km\,s$^{-1}$.  It is also seen that the unstable kink mode is a super-Alfv\'enic MHD wave with a relatively small growth rate Im$(\omega) = 0.236 \times 10^{-3}$~s$^{-1}$ at $k_z a = 1.743$---the maximum of the growth rate curve [see the cross point in \textbf{(b)} of Fig.~\ref{fig:fig2}] compared to the wave angular frequency Re$(\omega) = 175 \times 10^{-3}$~s$^{-1}$ [see the cross point in \textbf{(a)} of Fig.~\ref{fig:fig2}].  Both values of Re($\omega$) and Im($\omega$) have been evaluated in assuming that the tube radius $a = 1000$~km.  The corresponding wavelength of the kink mode is $\lambda_\mathrm{KH} = 3\,600$~km and its phase velocity is $100$~km\,s$^{-1}$.

We note, that at static magnetic flux tube ($\vec{v}_0 = 0$) the numerical code for finding the solution to Eq.~(\ref{eq:dispeqcom}) recovers the value of the normalized kink speed up to three places behind the decimal point.  It is intriguing to see how the KHI characteristics will change when we assume that the solar wind is an incompressible medium and its environment is a cool plasma.  This
\begin{figure}[!ht]
  \centering
\subfigure{\includegraphics[width = 3.3in]{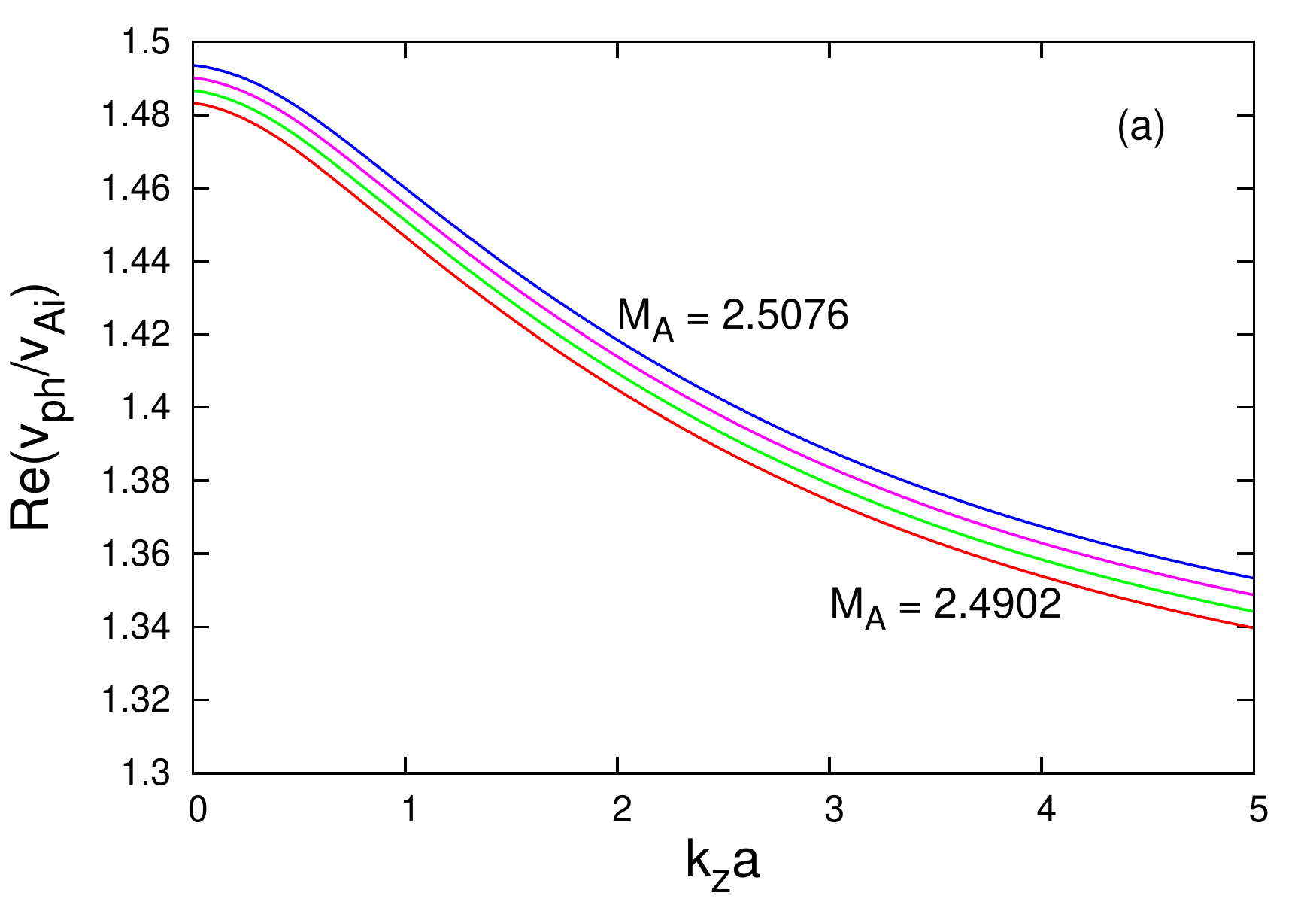}} \\
\subfigure{\includegraphics[width = 3.3in]{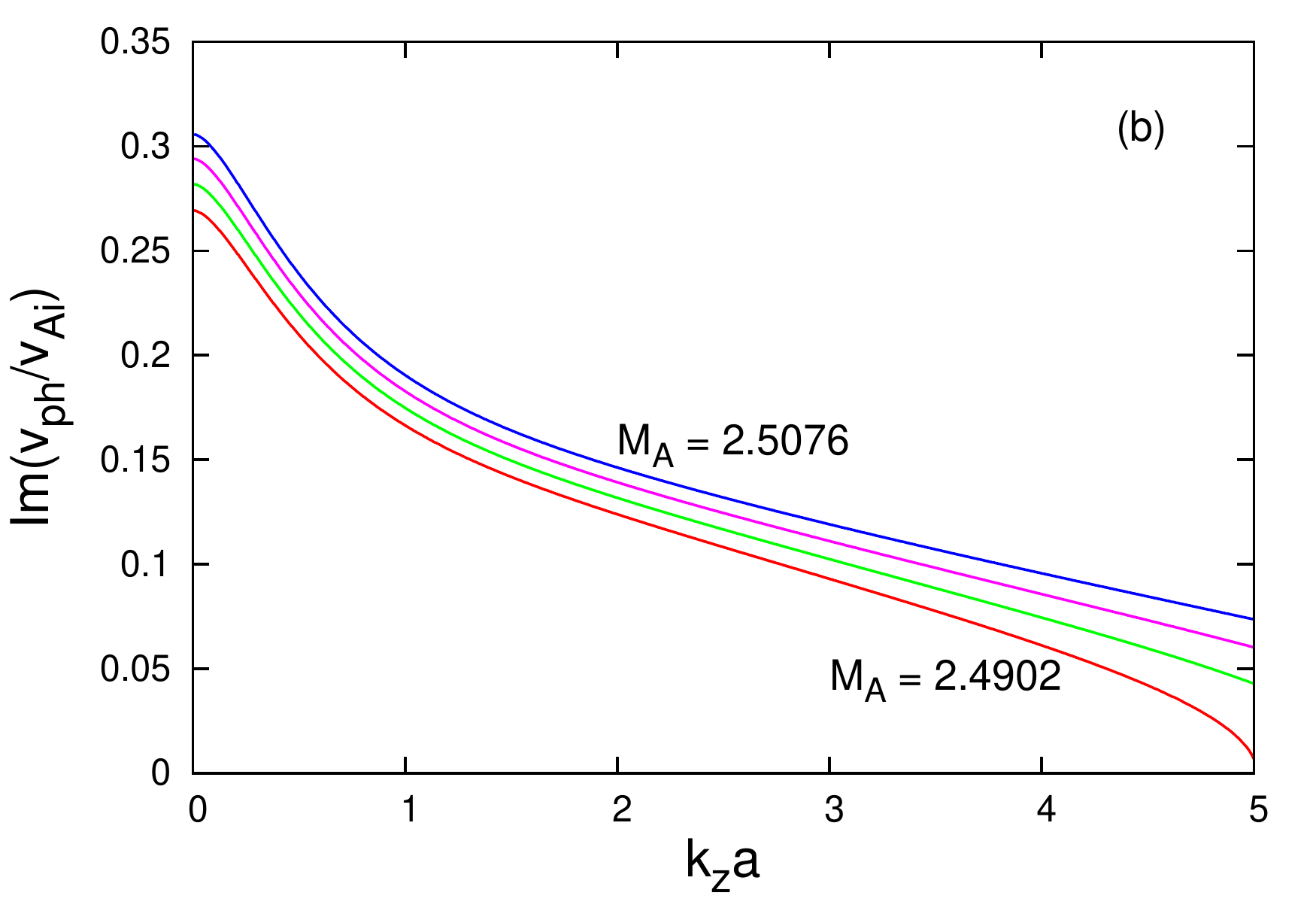}}
  \caption{\textbf{(a)} Dispersion curves of the kink mode ($m = 1$) propagating along an incompressible flowing magnetic flux tube surrounded by a cool medium at $\eta = 0.679$ and $b = 1.177$ for four Alfv\'en Mach numbers equal to $2.4902$, $2.496$, $2.5018$, and $2.5076$, respectively.  \textbf{(b)} Growth rates of the unstable kink mode at the same input parameters.  The marginally dispersion and growth rate curves are obtained at $M_\mathrm{A} = 2.4902$ and are plotted in red color.}
   \label{fig:fig3}
\end{figure}
\begin{figure}[!ht]
  \centering
\subfigure{\includegraphics[width = 3.3in]{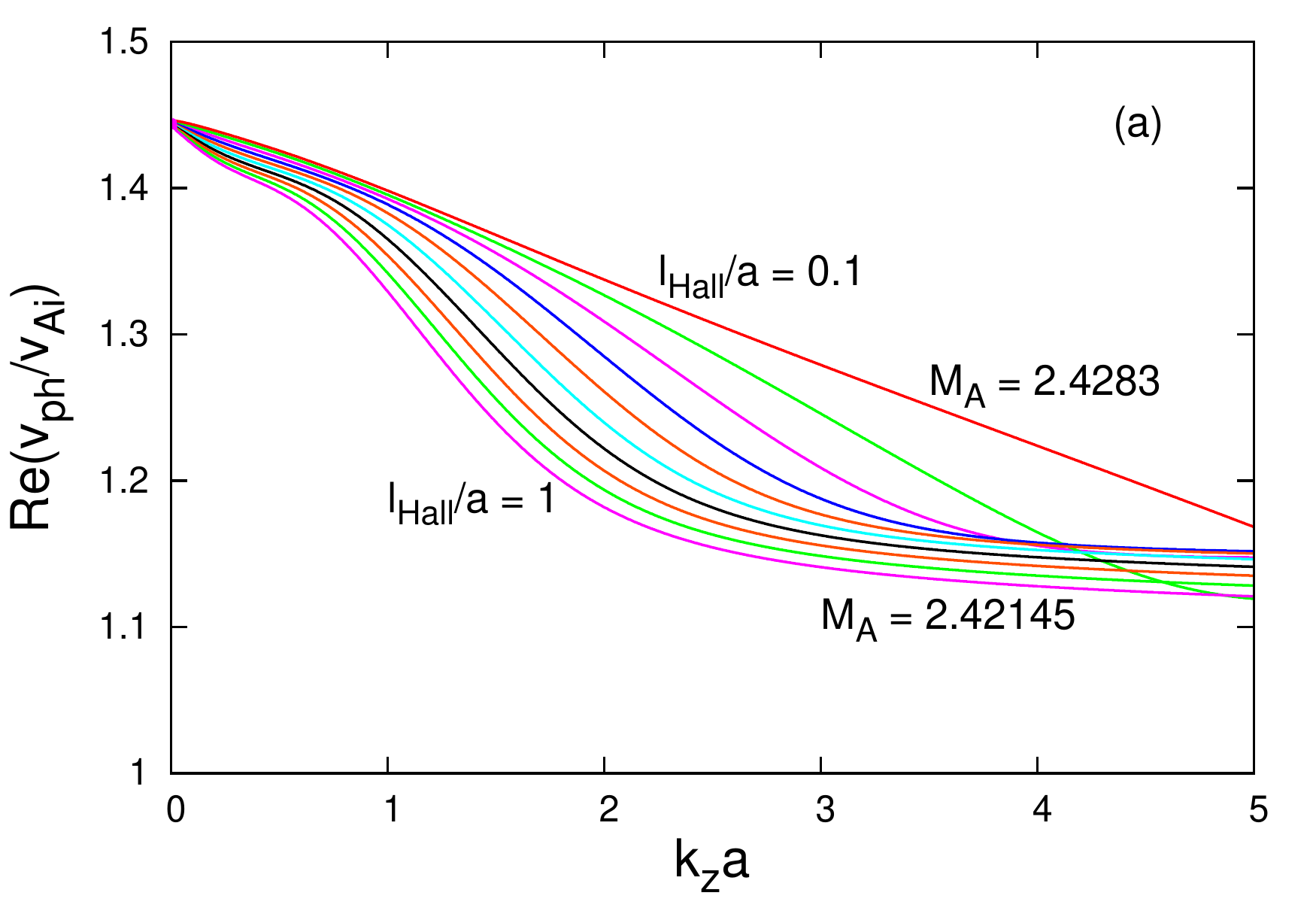}} \\
\subfigure{\includegraphics[width = 3.3in]{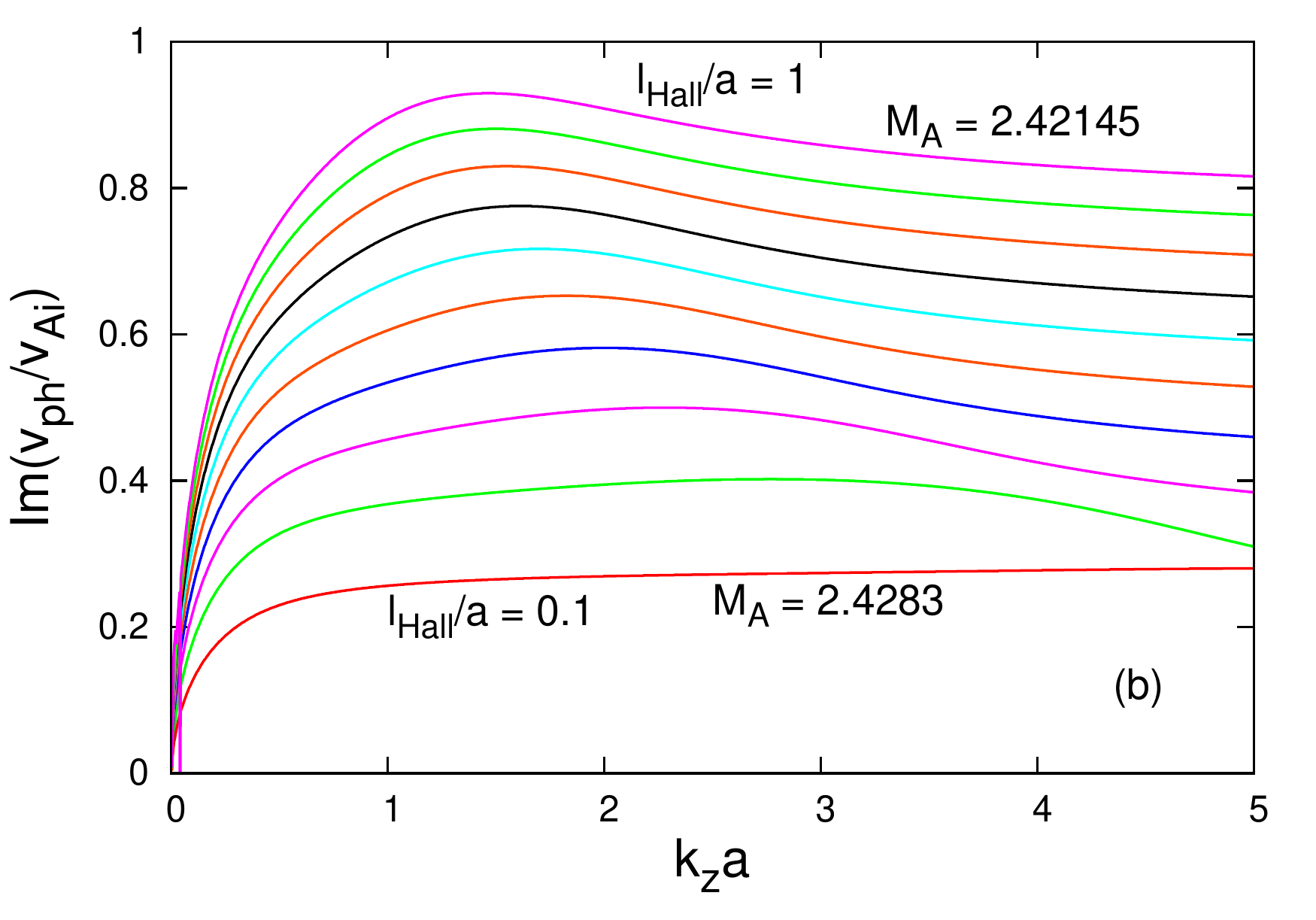}}
  \caption{\textbf{(a)} Dispersion curves of the kink mode ($m = 1$) propagating along an incompressible flowing magnetic flux tube surrounded by a cool medium in the framework of Hall-MHD at $\eta = 0.679$, $b = 1.177$, and $l_\mathrm{Hall}/a = 0.1,0.2,\mbox{\ldots},0.9,1$ for ten  Alfv\'en Mach numbers equal to $2.4283$, $2.4275$, $2.42675$, $2.426$, $2.4252$, $2.4245$, $2.42375$, $2.423$, $2.4222$, and $2.42145$, respectively.  \textbf{(b)} Growth rates of the unstable kink mode at the same input parameters.}
   \label{fig:fig4}
\end{figure}
means to find the solutions (in complex variables) to Eq.~(\ref{eq:dispeqmix}).  The input parameters now are: $m = 1$, $\eta = 0.679$, $b = 1.177$, and like before, $M_\mathrm{A}$ is a running parameter.  The dispersion curve and the growth rate of the kink mode are now pictured in Fig.~\ref{fig:fig3}.  The shape of both marginally curves (in red
color) is different of that of the similar curves plotted in Fig.~\ref{fig:fig2}, but the threshold Alfv\'en Mach number is almost the same, namely equal to $2.4902$, which implies a critical solar wind flow velocity of ${\cong}174$~km\,s$^{-1}$.  One sees that the difference between the two critical flow velocities for the appearance of the KHI is only $4$~km\,s$^{-1}$.  There is, however, one distinctive issue associated with the growth rate curve of KHI in Fig.~\ref{fig:fig3}, namely the instability region on the $k_z a$-axis possesses an upper limit at $k_z a \approx 5.0$.  If we want a wider instability region, it is necessary to increase the threshold Alfv\'en Mach number.  Nevertheless, we can conclude that the incompressible solar wind--cool surrounding plasma configuration describes more or less satisfactorily the KHI characteristics of the MHD kink mode ($m = 1$).  This conclusion allows us to use the same configuration in studying the propagation properties of the MHD waves in the framework of the Hall-MHD.  In such a case, it is necessary to accordingly modify the wave dispersion relation~(\ref{eq:dispeqn}), which will be made in the next subsection.

\subsection{Kelvin--Helmholtz instability of the kink and the \textit{m} = 2, 3, 4 MHD modes}
\label{subsec:kinkKHI}
Prior to begin the investigation of how the Hall current will affect the propagation and instability characteristics of the kink ($m = 1$) and a few higher MHD modes ($m = 2, 3, 4$), we have to slightly modify the wave dispersion relation~(\ref{eq:dispeqn}).  The modification concerns the wave attenuation coefficient of the solar wind cool environment, notably the $k_z$ in the arguments of $K_m$ and its derivative $K_m^\prime$ has to be replaced by $\kappa_\mathrm{e} = k_z(1 - \omega^2/k_z^2 v_\mathrm{e}^2)^{1/2}$.  Thus, the modified wave dispersion relation of Hall-MHD modes is:
\begin{eqnarray}
\label{eq:mixdispeqnkh}
    \frac{\rho_\mathrm{e}}{\rho_\mathrm{i}}\left( \omega^2 - k_z^2 v_\mathrm{Ae}^2 \right)Z_\mathrm{e}\left( k_z \frac{I_m^{\prime}(k_z a)}{I_m(k_z a)} - \frac{\varepsilon_\mathrm{i}}{1 - C_\mathrm{i}} \frac{m}{a} \right) \nonumber \\
    \nonumber \\
    {}-\left[ \left( \omega - k_z v_0 \right)^2 - k_z^2 v_\mathrm{Ai}^2 \right] \nonumber \\
    \nonumber \\
    {}\times Z_\mathrm{i}\left( \kappa_\mathrm{e} \frac{K_m^{\prime}(\kappa_\mathrm{e} a)}{K_m(\kappa_\mathrm{e} a)} - \frac{\varepsilon_\mathrm{e}}{1 - C_\mathrm{e}} \frac{m}{a} \right) = 0.
\end{eqnarray}

The input parameters for solving the above dispersion relation are the same as for finding the solutions to Eq.~(\ref{eq:dispeqcom}).  In addition, to explore how the Hall term affects the propagation and instability characteristics of the studied MHD modes, we introduce the parameter $l_\mathrm{Hall}/a$, whose values will vary from $0$ to $1$ with a step of $0.1$.  Thus, with $m = 1$, $\eta = 0.679$, $b = 1.177$, and $l_\mathrm{Hall}/a = 0.1,0.2,\mbox{\ldots},0.9,1$ we obtain the families of dispersion and growth rate curves shown in Fig.~\ref{fig:fig4}.  Here, we have two
striking things: (i) with the increase in the scale parameter, $l_\mathrm{Hall}/a$, the threshold Alfv\'en Mach number for arising of KHI gradually becomes lower; (ii) irrespective of the fact that the solar wind environment is a cool medium, the shape of the growth rate curves is not similar to that seen in the right panel of Fig.~\ref{fig:fig3}, that is, here there is no upper $k_z a$-limit.  The unstable kink mode ($m = 1$) is a super-Alfv\'enic wave [see \textbf{(a)} in Fig.~\ref{fig:fig4}] and the lowest critical jet velocity at which the mode becomes unstable is equal to $169.5$~km\,s$^{-1}$.  This happens at $l_\mathrm{Hall}/a = 1$.  For comparison, at $l_\mathrm{Hall}/a = 0.1$ the critical speed for the KHI onset is $170$~km\,s$^{-1}$, that is, the difference is rather small---only $0.5$~km\,s$^{-1}$.  Comparing with the value of $174$~km\,s$^{-1}$, found at exploring the KHI in the same configuration, but in the limit of standard ideal magnetohydrodynamics, we can conclude that the Hall term slightly diminishes the critical flow velocity of the solar wind for the emergence of KHI of the kink ($m = 1$) mode.

\begin{figure}[!hb]
  \centering
\subfigure{\includegraphics[width = 3.3in]{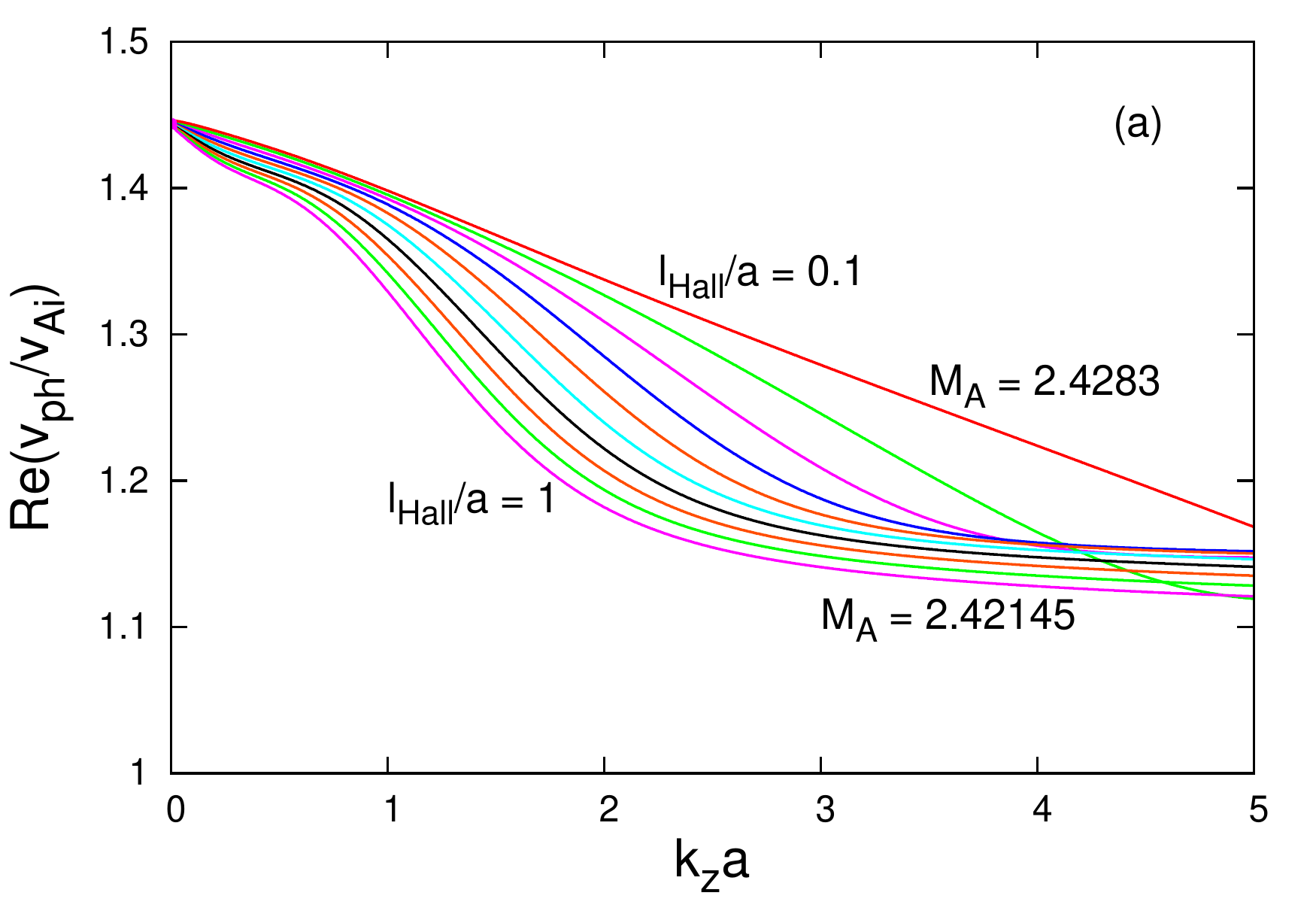}} \\
\subfigure{\includegraphics[width = 3.3in]{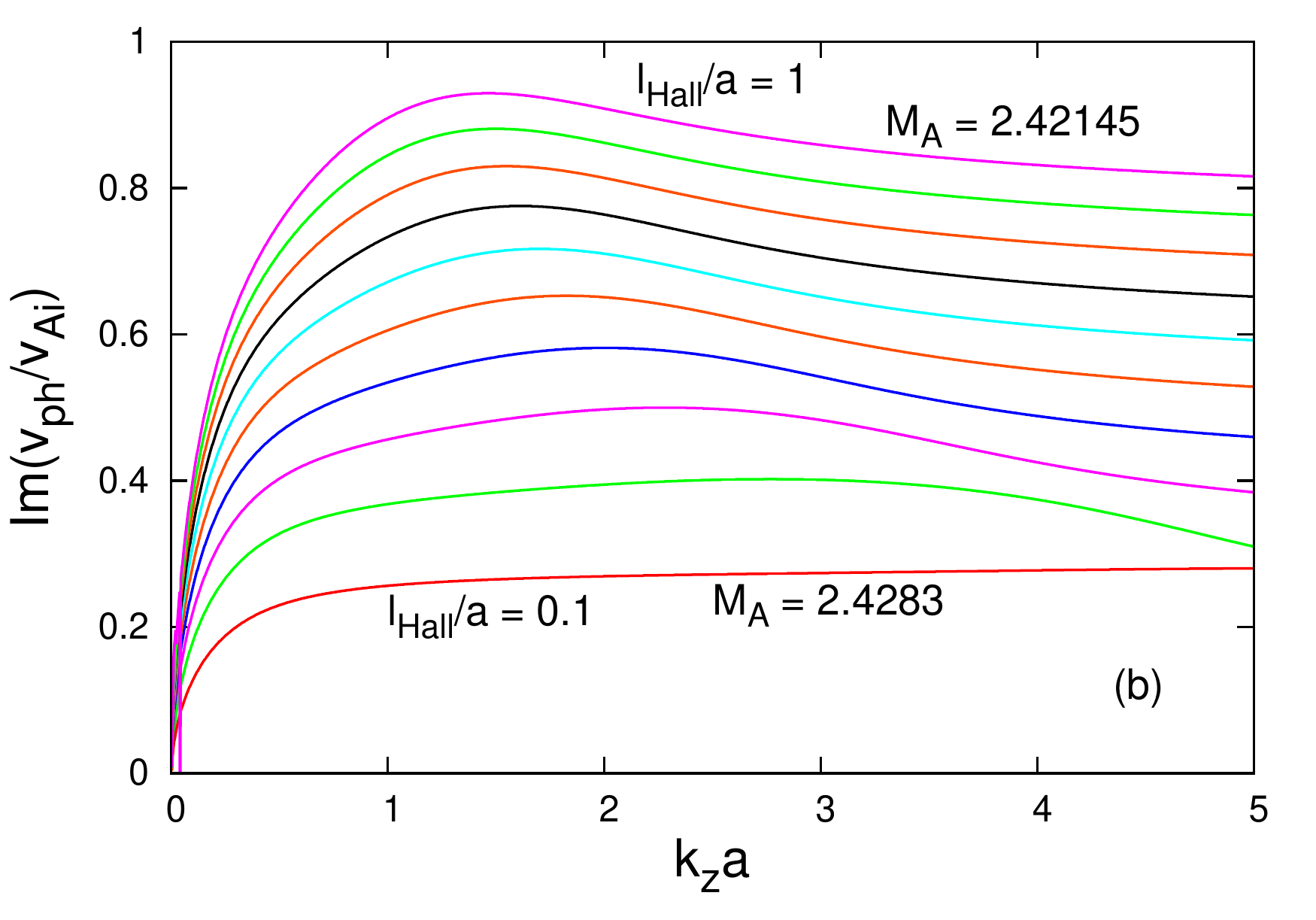}}
  \caption{The same as in Fig.~\ref{fig:fig4}, but with $b = 1$ and Alfv\'en Mach numbers $2.2233$, $2.2227$, $2.222$, $2.2214$, $2.219265$, $2.218289$, $2.217375$, $2.21645$, $2.215525$, and $2.2146$, respectively.}
   \label{fig:fig5}
\end{figure}

Since the magnetic field ratio, $b$, is close to $1$, we are tempted to see what will happen if we consider that both magnetic fields (internal and external) are the same, like in \citet{Zhelyazkov2010} and \citet{Zhelyazkov2018}.  Thus, performing the numerical calculations for finding the solutions to Eq.~(\ref{eq:mixdispeqnkh}) with the same input parameters, but with $b = 1$, we get the picture displayed below (see Fig.~\ref{fig:fig5}).  Not surprisingly, both figures, \ref{fig:fig4} and \ref{fig:fig5}, are very similar.  The only difference is that with equal magnetic fields, $b = 1$, the threshold Alfv\'en Mach numbers are lower
and subsequently the critical flow speed of the solar wind for the KHI startup of the kink mode ($m = 1$) is, in average, equal to $155.3$~km\,s$^{-1}$.

In studying the Hall current effect on the propagation and instability properties of the higher MHD modes ($m = 2, 3, 4$), along with the ten values of the scale parameter $l_\mathrm{Hall}/a$, we have calculated also the marginally dispersion and growth rate curves at $l_\mathrm{Hall}/a = 0$, that is,
\begin{figure}[!ht]
  \centering
\subfigure{\includegraphics[width = 3.3in]{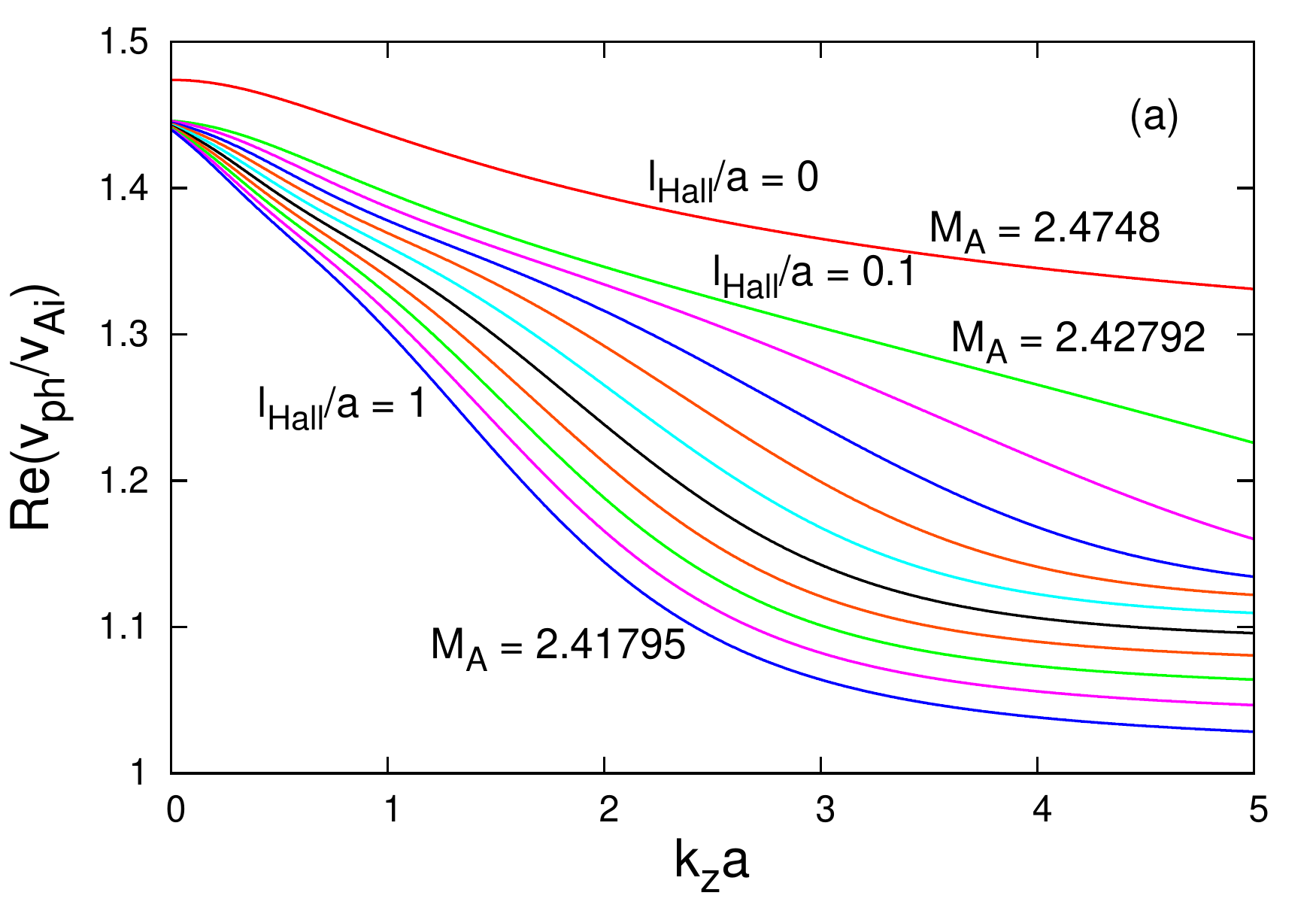}} \\
\subfigure{\includegraphics[width = 3.3in]{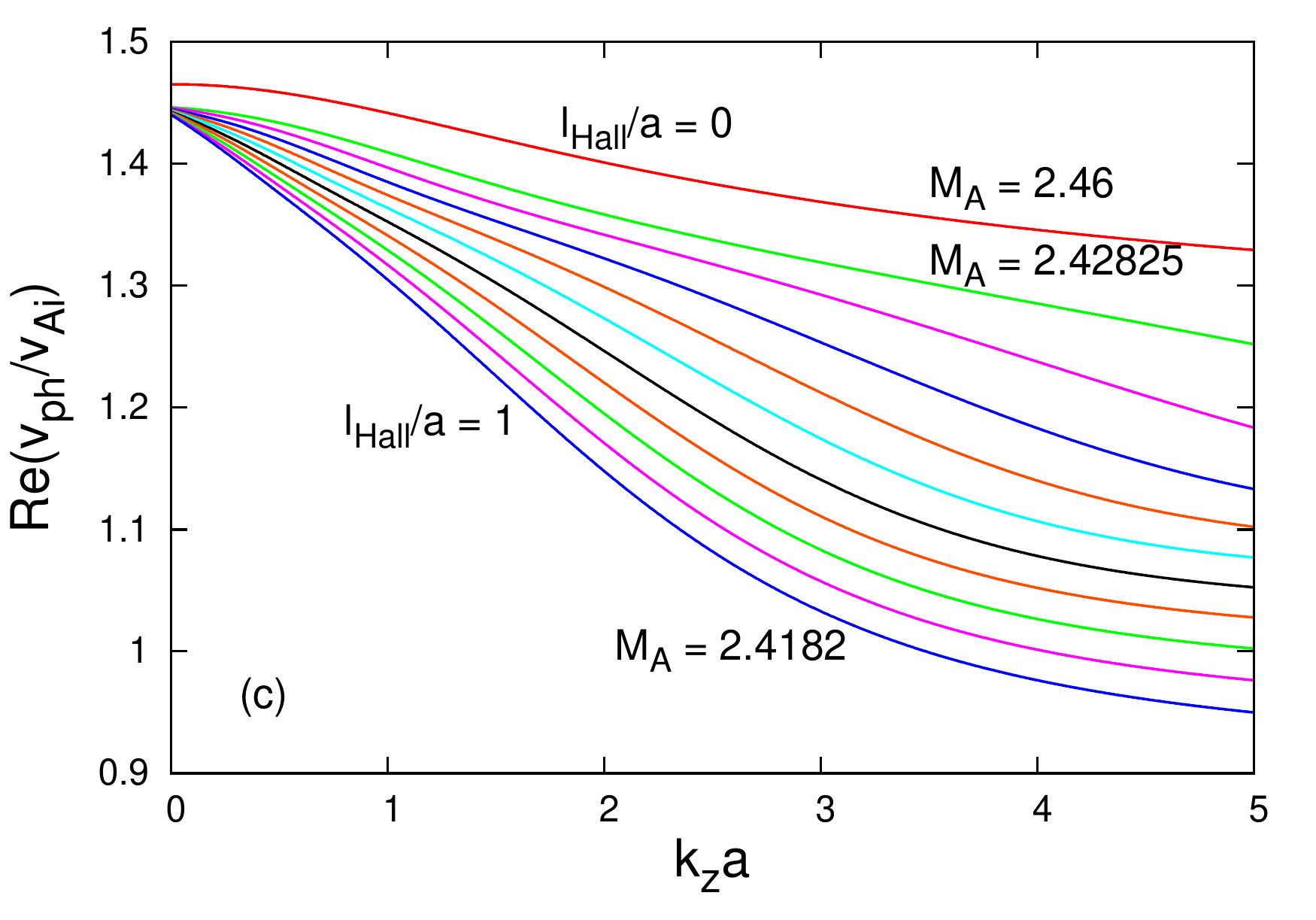}} \\
\subfigure{\includegraphics[width = 3.3in]{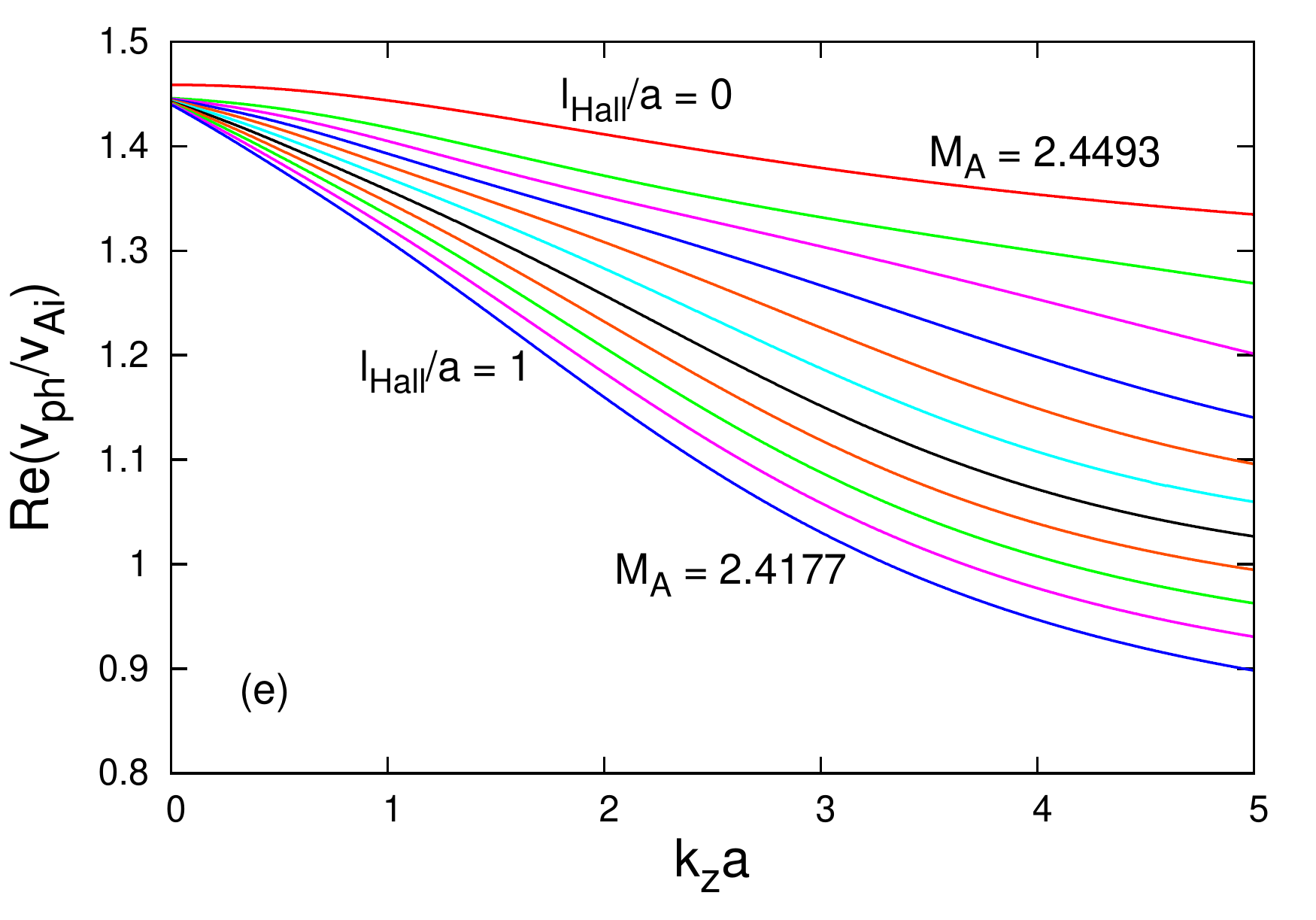}}
  \caption{\textbf{(a)}, \textbf{(c)}, \textbf{(e)} Dispersion curves of $m = 2,3,4$ MHD modes at the same input parameters as in Fig.~\ref{fig:fig4} plus the curves at $l_\mathrm{Hall}/a = 0$.}
   \label{fig:fig6}
\end{figure}
\begin{figure}[!ht]
  \centering
\subfigure{\includegraphics[width = 3.3in]{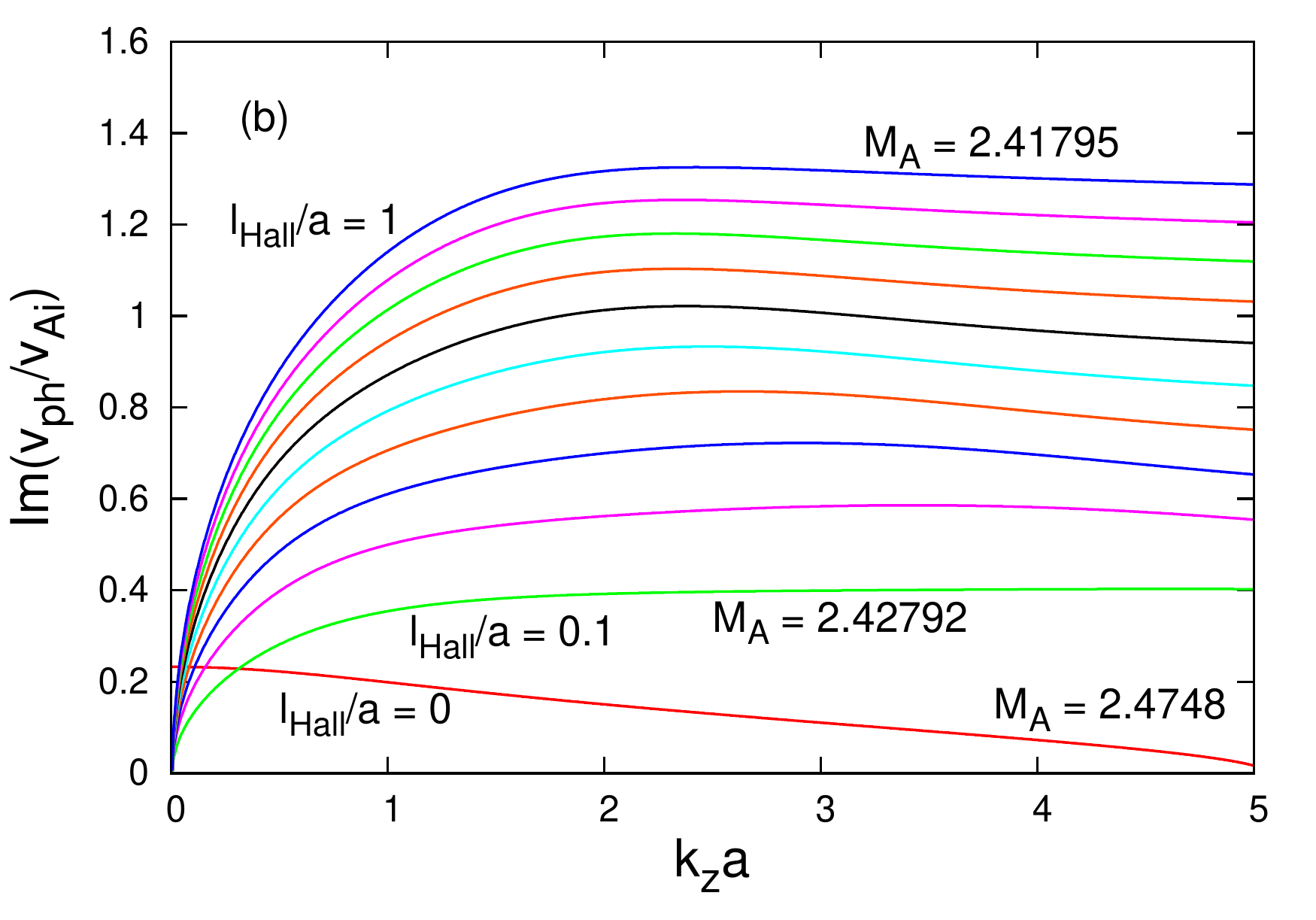}} \\
\subfigure{\includegraphics[width = 3.3in]{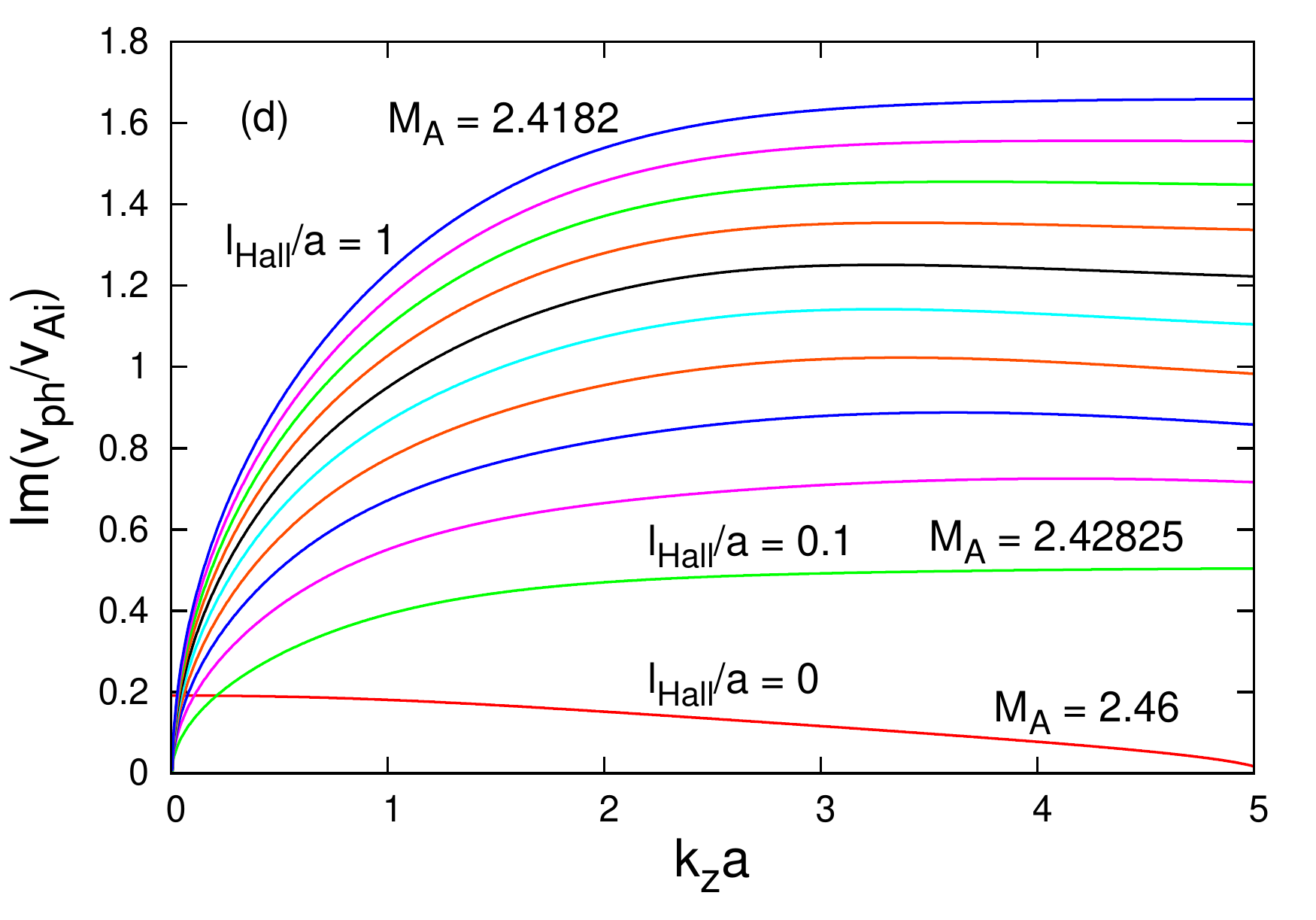}} \\
\subfigure{\includegraphics[width = 3.3in]{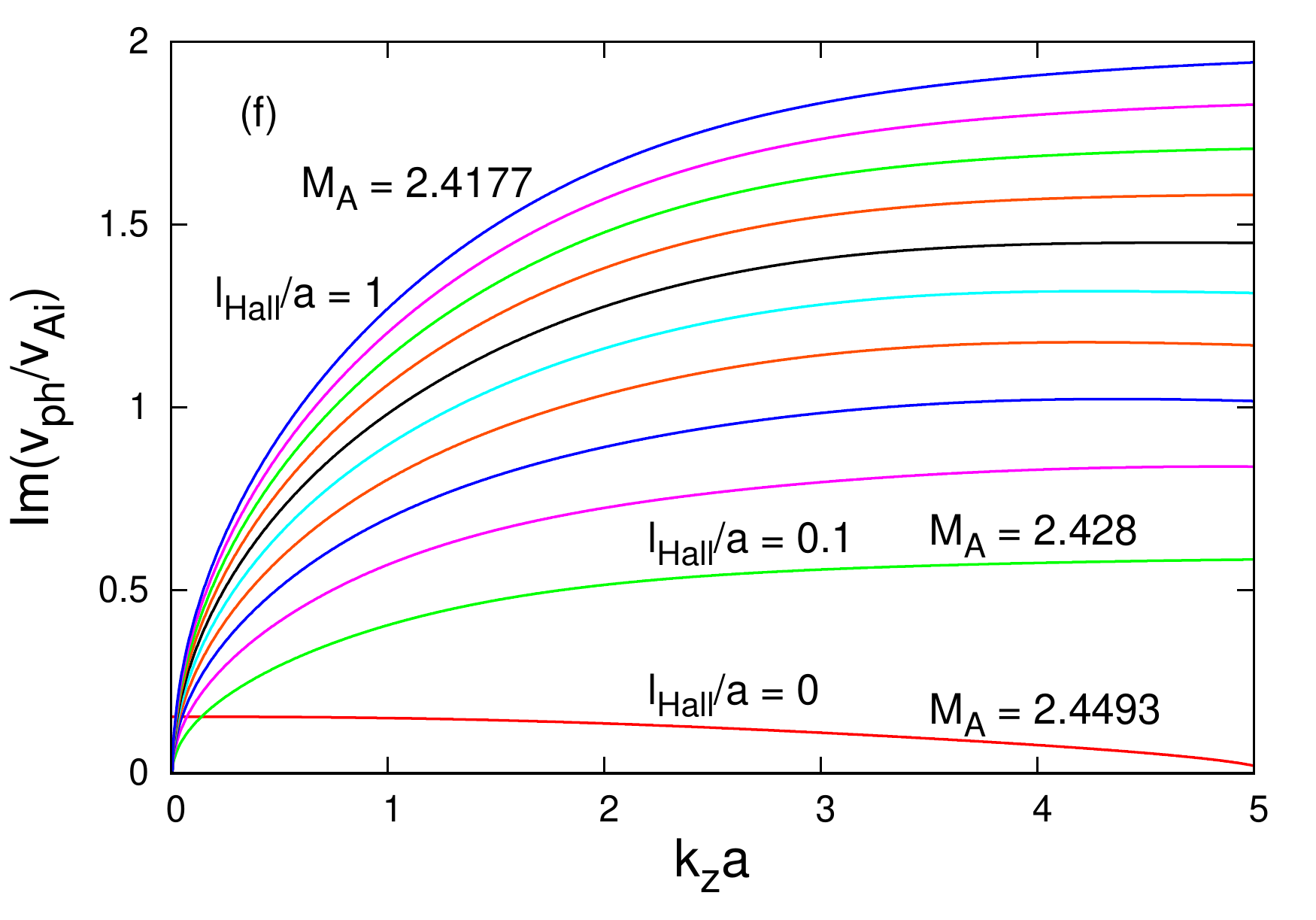}}
  \caption{\textbf{(b)}, \textbf{(d)}, \textbf{(f)} Growth rate curves of $m = 2,3,4$ MHD modes at the same input parameters as in Fig.~\ref{fig:fig4}.  The eleven threshold Alfv\'en Mach numbers for the $m = 2$ MHD mode are: $2.4748$, $2.42792$, $2.4275$, $2.42565$, $2.42445$, $2.423305$, $2.4222$, $2.42108$, $2.419925$, $2.418825$, $2.41795$; for the $m = 3$ mode: $2.46$, $2.42825$, $2.4275$, $2.4257$, $2.4245$, $2.4235$, $2.4222$, $2.421278$, $2.4199$, $2.41895$, $2.4182$; and for the $m =4$ mode: $2.4493$, $2.428$, $2.427$, $2.426$, $2.4249$, $2.4234$, $2.4222$, $2.42128$, $2.42$, $2.41889$, $24177$, respectively.}
   \label{fig:fig7}
\end{figure}
in the limit of the standard magnetohydrodynamics.  As seen from \textbf{(a)}, \textbf{(c)}, and \textbf{(e)} in Fig.~\ref{fig:fig6}, the threshold Alfv\'en Mach numbers are generally lower than
$M_\mathrm{A}^\mathrm{thr} = 2.4902$ for the kink ($m = 1$) MHD mode [see \textbf{(a)} in Fig.~\ref{fig:fig3}].  For the flute mode ($m = 2$), $M_\mathrm{A}^\mathrm{thr} = 2.4748$ which implies a critical solar-wind-flow speed of ${\cong}173$~km\,s$^{-1}$.  For the $m = 3$ mode we obtained $M_\mathrm{A}^\mathrm{thr} = 2.46$ (or equivalently a critical speed of ${\cong}172$~km\,s$^{-1}$), while for the $m = 4$ mode the numerics yield $M_\mathrm{A}^\mathrm{thr} = 2.4493$, which gives critical flow speed of $171.5$~km\,s$^{-1}$.  All these critical speeds are accessible for the slow solar wind.  It is rather interesting to observe that the marginally growth rate curves at $l_\mathrm{Hall}/a = 0$ [look at \textbf{(b)}, \textbf{(d)}, and \textbf{(f)} in figure~\ref{fig:fig7}] have the same pattern as the corresponding curve in \textbf{(b)} of Fig.~\ref{fig:fig3}.  As in the case of the kink ($m = 1$) mode, the Hall current diminishes $M_\mathrm{A}^\mathrm{thr}$ and its lowest value is $2.4177$ obtained for $m = 4$ and $l_\mathrm{Hall}/a = 1$, which means a critical solar-wind-flow speed of $169.3$~km\,s$^{-1}$.  Thus, one can conclude that the critical solar-wind-flow speed for the kink and higher MHD modes is around $170$~km\,s$^{-1}$ independently of the mode number, $m$, and the value of the scale parameter, $l_\mathrm{Hall}/a$.

\subsection{Kelvin--Helmholtz instability of the sausage MHD mode}
\label{subsec:sausageKHI}

\begin{figure}[!ht]
  \centering
\subfigure{\includegraphics[width = 3.3in]{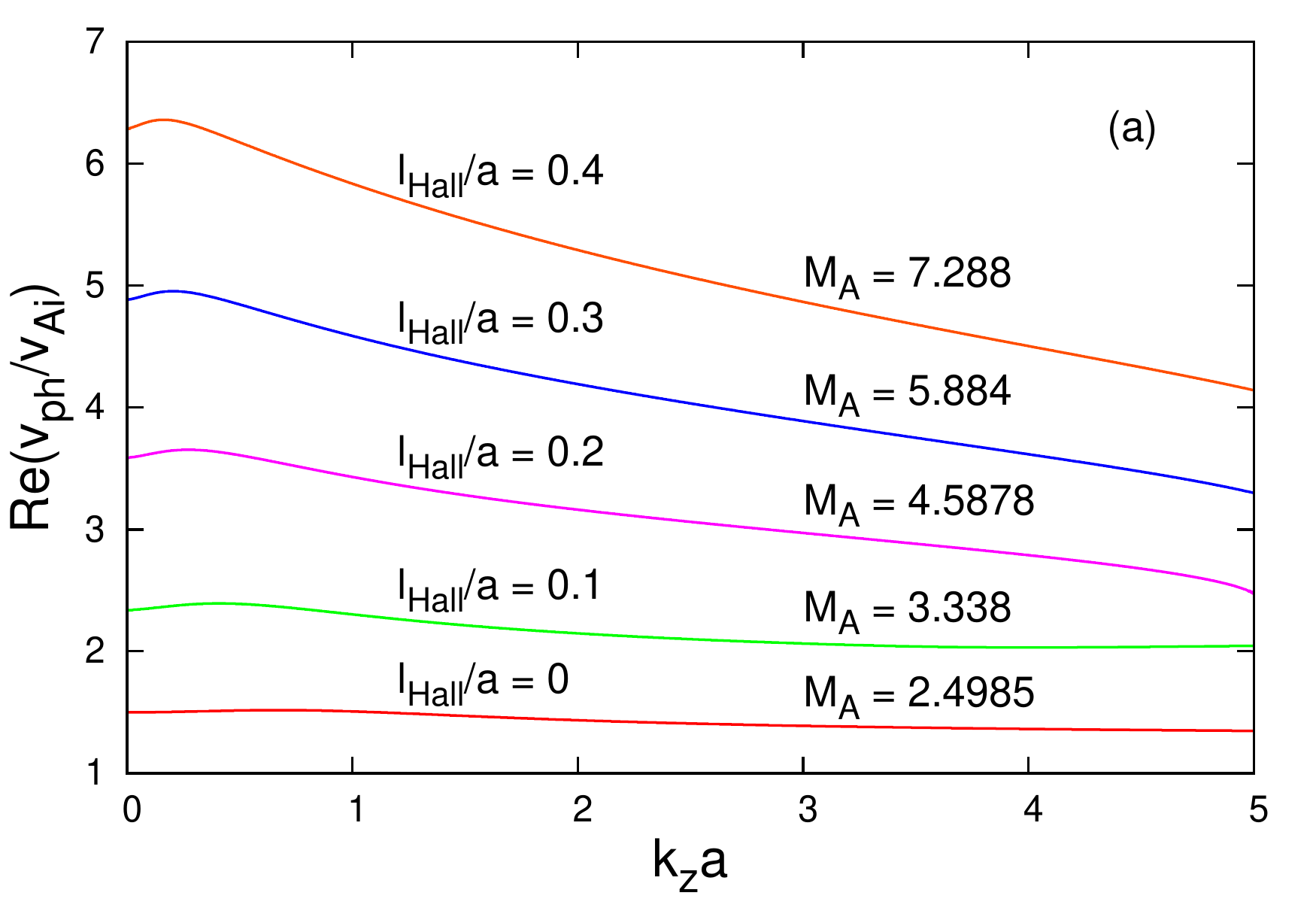}} \\
\subfigure{\includegraphics[width = 3.3in]{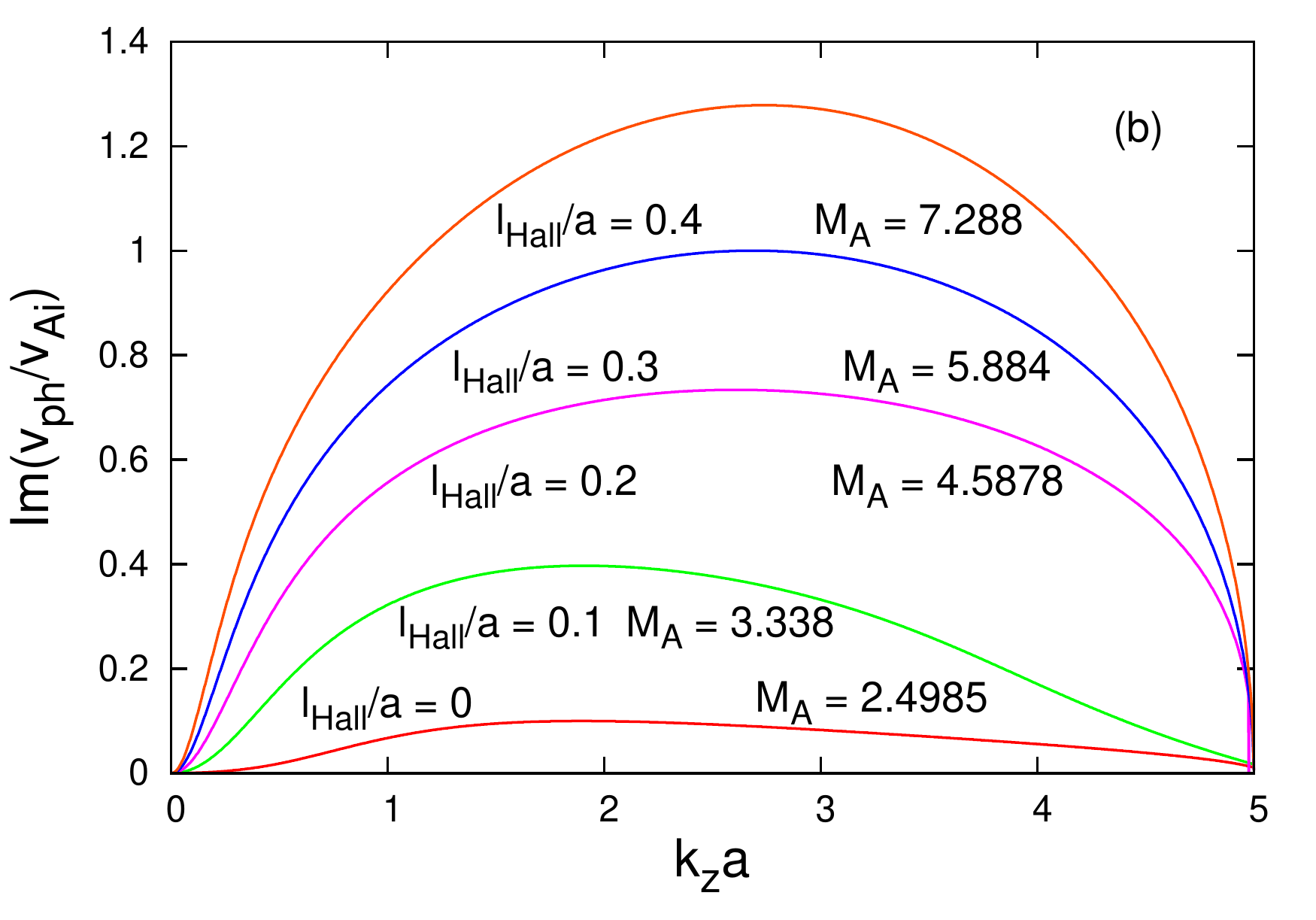}}
  \caption{\textbf{(a)} Dispersion curves of the sausage mode ($m = 0$) propagating along an incompressible flowing magnetic flux tube surrounded by a cool medium in the framework of Hall-MHD at $\eta = 0.679$, $b = 1.177$, and $l_\mathrm{Hall}/a = 0.1,0.2,0.3,0.4$ for four  Alfv\'en Mach numbers equal to $3.338, 4.5878, 5.884, 7.288$, respectively.  The red color dispersion curve has been computed with $l_\mathrm{Hall}/a = 0$, \emph{i.e.}, in the framework of standard MHD.  \textbf{(b)} Growth rates of the unstable sausage mode at the same input parameters.}
   \label{fig:fig8}
\end{figure}

In contrast to the kink ($m = 1$) and the high-mode ($m \geqslant 2$) MHD waves, the sausage ($m = 0$) mode is noticeably influenced by the Hall current.  There is a distinct difference in the wave response to the value of the scale parameter $l_\mathrm{Hall}/a$ of this mode, notably while the increasing in $l_\mathrm{Hall}/a$ the threshold value of Alfv\'en Mach number, $M_\mathrm{A}^\mathrm{thr}$, for KHI onset of the kink and the high-mode MHD waves diminishes, for the sausage wave the trend is opposite.  In the first case, as it was discussed in the previous subsection, the average critical solar-wind-flow speed is of the order of $170$~km\,s$^{-1}$ independently of the magnitude of $l_\mathrm{Hall}/a$.  The numerical solutions to the wave dispersion relation~(\ref{eq:mixdispeqnkh}) with $m = 0$ show a steep growing in $M_\mathrm{A}^\mathrm{thr}$, and subsequently in the critical flow velocity for the KHI occurrence.  The results of numerical computations with the same input parameters as those in Fig.~\ref{fig:fig4}, for four values of the scale parameter $l_\mathrm{Hall}/a$ equal to $0.1$, $0.2$, $0.3$, and $0.4$, respectively, are illustrated in Fig.~\ref{fig:fig8}.  For comparison, we
have also computed the marginally dispersion and growth rate curves with $l_\mathrm{Hall}/a = 0$, which are analogs of the red-colored curves in Fig.~\ref{fig:fig3}.  The critical flow speed for rising the KHI of the sausage mode at $l_\mathrm{Hall}/a = 0$ is ${\cong}175$~km\,s$^{-1}$, which is accessible in the slow solar wind.  But even at relatively small Hall scale parameters like $0.3$ and $0.4$, the critical flow speeds for a KHI emerging are much higher, equal respectively to ${\cong}412$ and $510$~km\,s$^{-1}$, both being inaccessible in the studied case.  We can claim that the Hall current has a stabilizing effect on the KHI occurrence of the sausage mode.  It is worth noticing that the shape of growth rates curves [see Fig.~\ref{fig:fig8}(b)] is typical for an incompressible jet--cool environment configuration, that is, a wider instability region than that shown in Fig.~\ref{fig:fig8} requires higher threshold Alvf\'en Mach numbers and higher flow speeds.

\section{Conclusion and outlook}
\label{sec:conclusion}
In this article, we have studied the propagation and instability characteristics of MHD modes propagating along a solar-wind flowing plasma in the frameworks of both ideal standard and Hall magnetohydrodynamics.  We model the jet as a moving cylindrical magnetic flux tube (with radius $a$) of homogeneous density $\rho_\mathrm{i}$ immersed in a homogeneous magnetic field $\vec{B}_\mathrm{i}$ and surrounded by static plasma with homogeneous density $\rho_\mathrm{e}$ and constant magnetic field $\vec{B}_\mathrm{e}$.  In finding the appropriate wave dispersion equation in the second case of Hall-MHD, we critically reconsider the basic MHD equations which govern the dynamics of solar wind and %
\begin{figure}[!ht]
  \centering
\subfigure{\includegraphics[width = 3.3in]{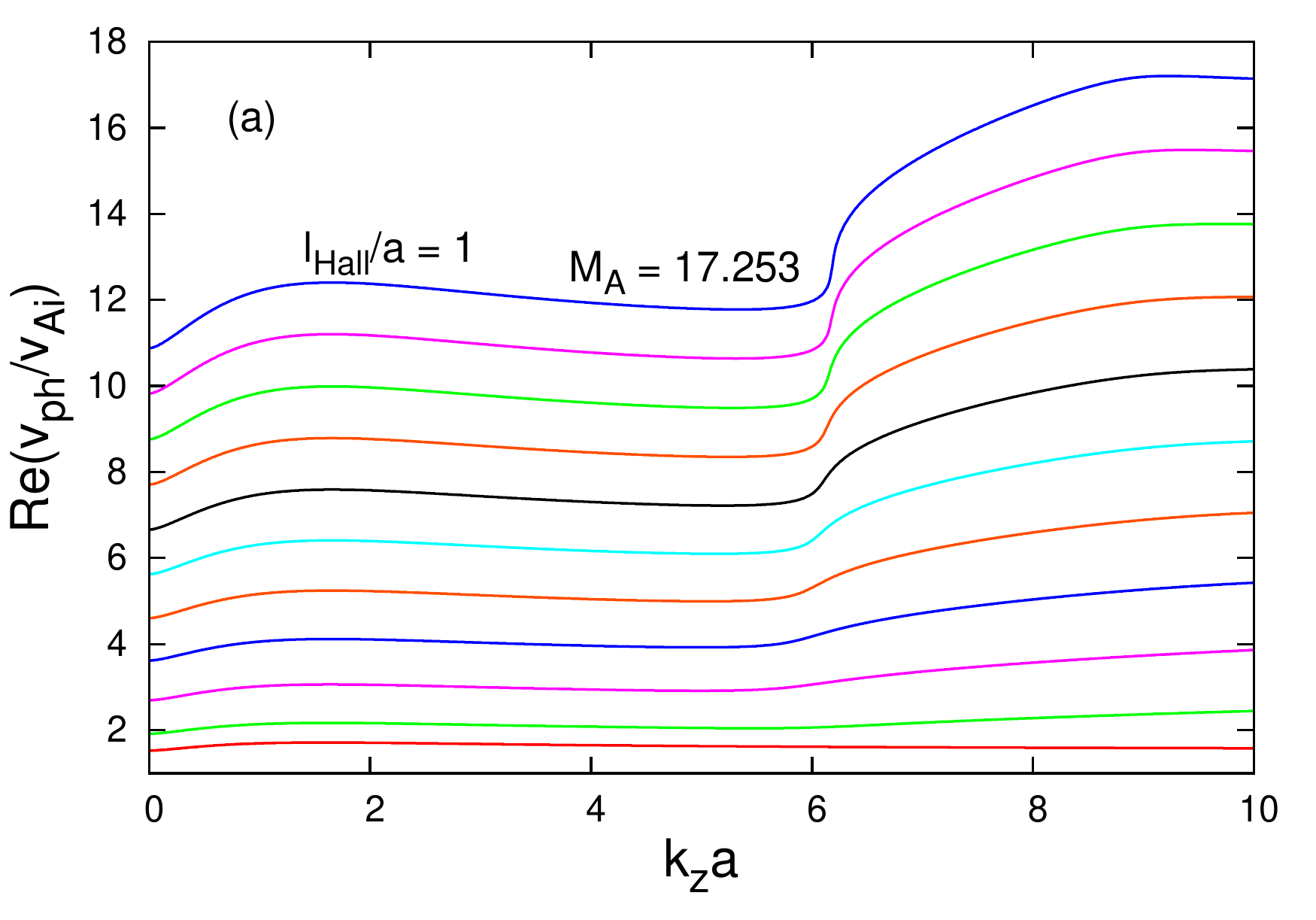}} \\
\subfigure{\includegraphics[width = 3.3in]{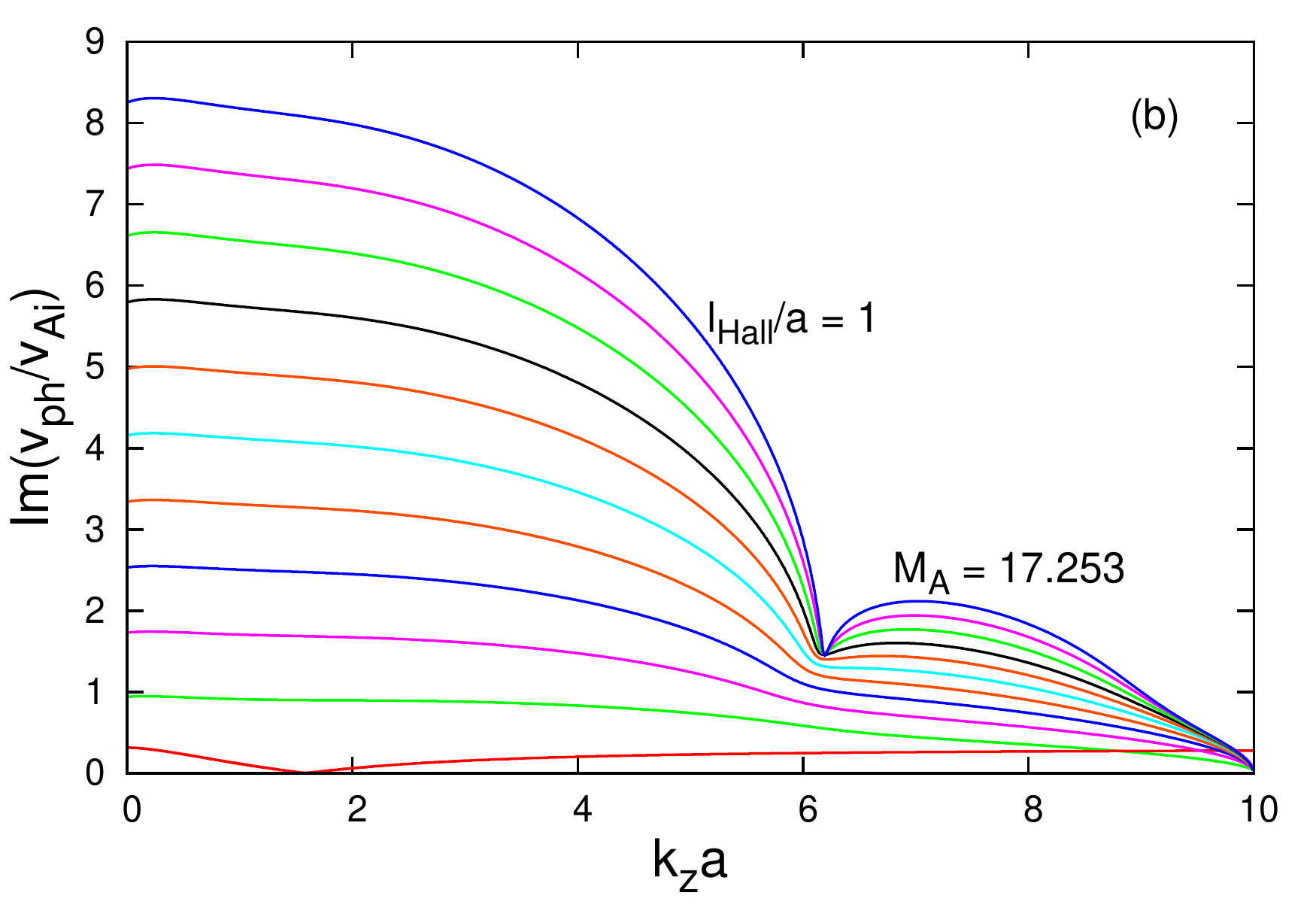}}
  \caption{\textbf{(a)} Dispersion curves of the kink mode ($m = 1$) propagating along an incompressible flowing magnetic flux tube surrounded by an incompressible medium in the framework of Hall-MHD obtained from Eq.~(16) at $\eta = 0.586$, $b = 1$, and $l_\mathrm{Hall}/a = 0,0.1,0.2,\mbox{\ldots},0.9,1$ for eleven Alfv\'en Mach numbers equal to $2.4161$, $3.035$, $4.272$, $5.7365$, $7.3$, $9.92$, $10.563$, $12.225$, $13.895$, $15.583$, $17.253$, respectively.  \textbf{(b)} Growth rates of the unstable kink mode at the same input parameters.}
   \label{fig:fig9}
\end{figure}
\begin{figure}[!ht]
  \centering
\subfigure{\includegraphics[width = 3.3in]{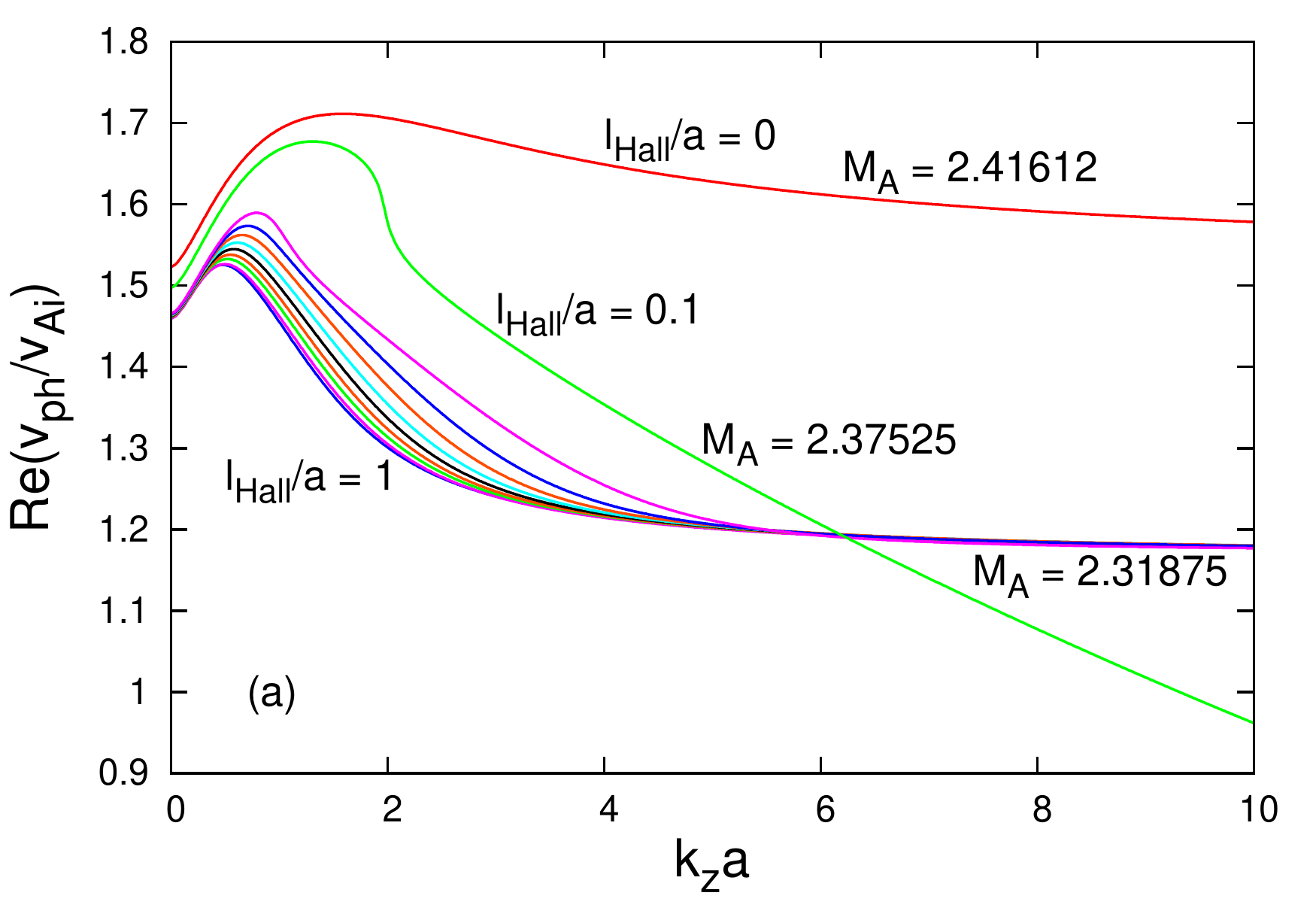}} \\
\subfigure{\includegraphics[width = 3.3in]{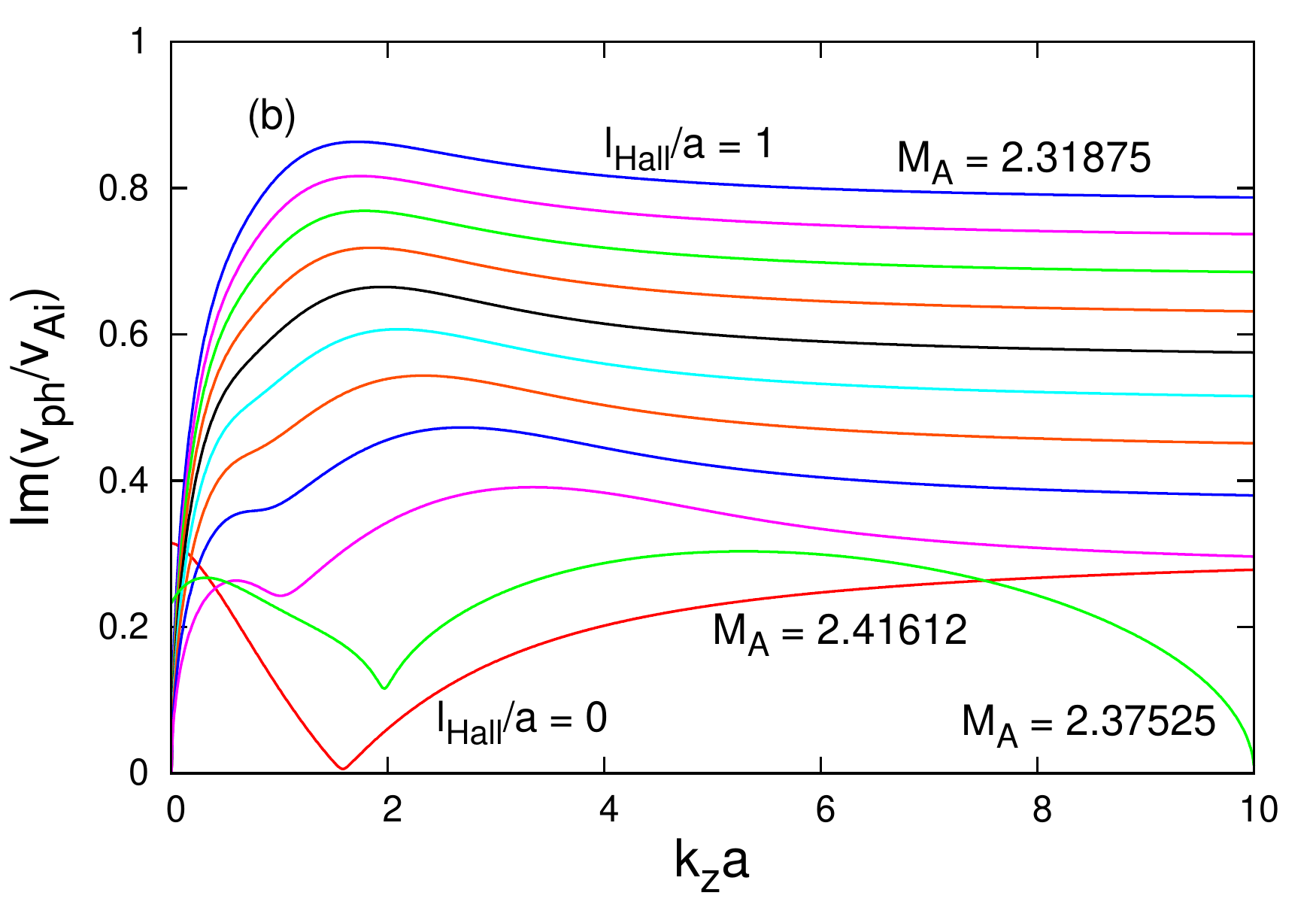}}
  \caption{\textbf{(a)} The same input parameters as in Fig.~\ref{fig:fig9}(a), but for solving Eq.~(\ref{eq:dispeqn}) with the following Alfv\'en Mach numbers: $2.41612$, $2.3752$, $2.325$, $2.3235$, $2.3225$, $2.321$, $2.3195$, $2.318$, $2.3175$, $2.315$, and $2.31875$.  \textbf{(b)}  Growth rates of the unstable kink mode at the same input parameters.}
   \label{fig:fig10}
\end{figure}
its environment plasmas. More specifically, bearing in mind the chosen plasma and magnetic field parameters of the slow solar wind, we treat the jet plasma (with plasma beta grater than $1$) as an incompressible medium whilst the surrounding plasma possessing a lower (less than $1$) plasma beta is considered as a cool medium.  As we have mentioned in Section~1, the major drawback in \citet{Zhelyazkov2010}, where there was investigated the similar problem, is the circumstance that when considering the jet plasma and its environment as incompressible media the thermal pressure in the governing momentum equation of the Hall-MHD was neglected.  Here, we kept that term which finally
yielded a new dispersion Eq.~(\ref{eq:mixdispeqnkh}) which alongside the propagation of the kink ($m = 1)$ MHD mode describes also the propagation of higher-mode ($m \geqslant 2$) MHD waves.  The solutions to the derived dispersion Eq.~(\ref{eq:mixdispeqnkh}) show that the Hall term does not change significantly the value of the threshold Alfv\'en Mach number, $M_\mathrm{A}^\mathrm{thr}$, for instability onset compared with that obtained in the framework of standard MHD.  Furthermore, it (the Hall current) stimulates the occurrence of KHI due to a diminishing of $M_\mathrm{A}^\mathrm{thr}$ and accordingly to the critical flow velocity yielding around $170$~km\,s$^{-1}$ for the kink and higher MHD modes.  In sharp contrast, the wave dispersion relation~(16) in \citealp{Zhelyazkov2010} exhibits just the opposite trend---the Hall current increases the threshold Alfv\'en Mach number and the corresponding critical jet speed.  These distinctive trends are better illustrated by comparing the results obtained from the solutions to the wave dispersion relations (16) and (\ref{eq:dispeqn}) (both applicable for incompressible plasmas in the solar wind and its environment) with the same input data, namely $\eta = 0.586$, $b = 1$ (equal magnetic fields in the two media), and eleven values of the Hall scale parameter, $l_\mathrm{Hall}/a$.  In \textbf{(a)} and \textbf{(b)} of Fig.~\ref{fig:fig9} one sees the plots deduced from the solutions to Eq.~(16) in \citealp{Zhelyazkov2010} and in \textbf{(a)} and \textbf{(b)} of Fig.~\ref{fig:fig10}---those from Eq.~(\ref{eq:dispeqn}).  The red curves in all plots are the marginally curves for emerging the KHI of the kink ($m = 1$) mode in the jet.  Not surprisingly, they are identical for the two approaches.  The influence of the Hall current on the propagation and instability characteristics of that mode are, however, completely different.  If we choose a Hall scale parameter $l_\mathrm{Hall}/a = 0.4$---the value that was used in \citet{Zhelyazkov2010}---from the orange curve in Fig~\ref{fig:fig9}(a), with $v_\mathrm{Ai} = 70$~km\,s$^{-1}$ and $M_\mathrm{A}^\mathrm{thr} = 7.3$, we find that the critical jet speed for KHI onset is $v_{0}^\mathrm{cr} = 511$~km\,s$^{-1}$, a value too high for the slow solar wind.  From the analogous curve in Fig.~\ref{fig:fig10}(a) we obtain $v_{0}^\mathrm{cr} = 162.5$~km\,s$^{-1}$, which, obviously, is an affordable solar-wind speed.  In using the erroneous approach (neglecting the thermal pressure in the Hall-MHD momentum equation), one concludes that at $l_\mathrm{Hall}/a = 0.4$ the Hall current suppresses the instability emerging, while the correct treatment of the problem yields that the Hall current even stimulates the KHI occurrence ($162.4$~km\,s$^{-1}$ \emph{vs.}\ ${\cong}169$~km\,s$^{-1}$ at $l_\mathrm{Hall}/a = 0)$.  It is worth underlying also the difference in the shapes of growth rate curves pictured in Fig.~\ref{fig:fig9}(b) and Fig.~\ref{fig:fig10}(b): the imaginary solutions to Eq.~(16) in \citealp{Zhelyazkov2010} yield untypical growth rate curves which are limited on the $k_z a$-axis, while the instability range, obtained from the solutions to Eq.~(\ref{eq:dispeqn}), is practically unlimited.

The picture for the sausage ($m = 0$) mode is also entirely different---in using Eq.~(16) in \citealp{Zhelyazkov2010} with $m = 0$ one concludes that the sausage mode is not influenced by the Hall current, by contrast to the new approach according to which the Hall current represses the KHI rising at $l_\mathrm{Hall}/a > 0.3$.

To sum up, our reconsidered approach in modeling of the KHI in the framework of the ideal Hall-MHD yields dispersion and growth rate curves of the unstable kink ($m = 1$) mode, which in many ways are similar to those obtained in the opposite case when both the jet and its environment are treated as cool media---compare, for example, our Fig.~\ref{fig:fig4} and Figure~5 in \citealp{Zhelyazkov2018}.  The real challenge in the modeling the KHI of the MHD modes in cylindrical solar-wind geometry according to the seminal article by \citet{Lighthill1960} is to study how the plasma compressibility in both media (the jet and its environment) will change the propagation and instability properties of the unstable MHD modes due to the Hall term in the induction equation.  Another direction for improving this modeling is to take into account the twist of the magnetic field which is typical for the most jets in the solar atmosphere.

\acknowledgments
The work was supported by the Bulgarian Science Fund under project DNTS/INDIA 01/7.  We are deeply indebted to the reviewer for his very helpful and constructive comments.  The authors are also thankful to Dr.~Snezhana Yordanova for plotting one figure.

\section*{ORCID iD}
\vspace{1.5mm}
I.~Zhelyazkov {\color{blue}orcid.org/0000-0001-6320-7517}

\end{document}